\documentstyle[twocolumn,pre,aps,epsf]{revtex}
\begin{document}

\title{Friction in a solid lubricant film}

\author{O.\ M.\ Braun$^{1,}$\thanks{Electronic address: obraun@iop.kiev.ua}
and M.\ Peyrard$^{2}$}

\address{$^{1}$
Institute of Physics,
National Ukrainian Academy of Sciences,
03650 Kiev, Ukraine}

\address{$^{2}$
Laboratoire de Physique de l'Ecole Normale Sup\'{e}rieure de Lyon,\\
46 All\'{e}e d'Italie,
69364 Lyon C\'{e}dex 07, France}

%\date{\today}
\maketitle

\begin{abstract}

Molecular dynamics study of a thin (one to five layers)
lubricant film between two substrates in moving contact are performed
using Langevin equations with an external damping
coefficient depending on  distance and velocity of atoms relative
the substrates, motivated by microscopic configurations.
They show that the minimal friction coefficient is
obtained for the solid-sliding regime. A detailed analysis of the
results, the comparison with other microscopic modeling approaches of
friction, and the evaluation of quantities that can be compared to
experiments, such as the velocity of the transition from stick-slip to
smooth sliding, are used to discuss the relevance of the microscopic
simulations of friction.

\end{abstract}

\pacs{PACS numbers: 46.30.Pa; 81.40.Pd; 61.72.Hh; 46.10.+z}

%=========================================================================
\section{Introduction}

The problem of friction between two substrates 
which are in moving contact
is very important technologically as well as
very rich physically 
\cite{BT1986,P0,P00,Mate95,Gen,Gen2,Robbins2}.
Following the development of atomic force microscopy,
tribology has approached the microscopic level and
these studies are expanding
because nanotechnology is now building  devices
that are so miniaturized that they begin to probe 
the microscopic properties of the materials.
But the atomic friction microscope
has its own difficulties, in particular for
``high-speed friction'',
because it does not operate at high velocities.
This explains the interest of numerical simulations of friction
\cite{Gao} (see also the review paper
\cite{Robbins2}
and references therein).
Moreover molecular dynamics can provide detailed information on the
motion of individual atoms which help in the understanding of the
basic mechanisms of friction, a problem which is still open in spite
of the efforts devoted to it in the last few years.
Unfortunately, large-scale simulation of dynamical processes with
three-dimensional models and realistic interatomic potentials
are still too time consuming for large systems, mainly due to slow
relaxation processes in such systems.
Moreover the standard molecular dynamics methods
consider the atomic degrees of freedom only
and ignore the electronic ones,
that may be a too crude approximation
in modeling metal substrates.
Therefore, one has to use simplified models such as two-dimensional
\cite{Jeam,Persson,force}
or even one-dimensional \cite{BBR97,Urbakh-FK} ones.

An intermediate complexity model,
where a few atomic-layer film is placed between two rigid substrates,
so that the top and bottom lubricant layers may stick to the
substrates while leaving mobile atoms in between,
was analyzed in a series of papers 
\cite{HMR1999-MR2000b-MR2000a,TR1990a,SR1993,Robbins,SRC1996,HR2000}
and also in our previous publications for
commensurate \cite{BDP97} and
incommensurate \cite{BBR99} lubricant films.
Thermal equilibrium was achieved with the help of
Langevin equations with a random Gaussian force and
a viscous damping with a constant coefficient $\eta$.
The Langevin thermostat was applied either to all
degrees of freedom of the system \cite{BDP97,BBR99},
or only to the ones perpendicular to the sliding direction
\cite{HMR1999-MR2000b-MR2000a,TR1990a,SR1993,Robbins,SRC1996,HR2000}.
Owing to the relative simplicity of such models they allow
the simulation of friction for a broad range of parameters
and the study of general trends and laws of the phenomena.
The main results obtained in these studies are the following.
First, it was shown that
the static frictional force $f_s$ for the contact
of bare substrates is very small in most cases,
except in the rare situation when the two surfaces
are commensurate and perfectly aligned
\cite{Robbins2,HS1990,SJS1996}.
However, even a sub-monolayer of mobile atoms
between the surfaces leads to a finite $f_s$
which is proportional to the applied load
and is almost independent on the system parameters
\cite{HMR1999-MR2000b-MR2000a}.
Second, a thin lubricant film 
(with a thickness smaller than $\sim 10$ molecular diameters)
confined between two solids
is always layered and often solidified because
the confinement decreases the entropy of the film
and shifts the bulk melting transition
to higher temperatures and lower pressures
(see, e.g., \cite{Robbins2} and references therein).
Third, when an external force $f$ drives one of the substrates
and it starts to move,
the lubricant film melts
and its width increases by $\sim 10$\%
\cite{TR1990a,SR1993}.
After the shear-induced first-order melting transition,
the lubricant exhibits a
layer-over-layer sliding
with strong two-dimensional order 
within shearing planes of atoms, where
each layer moves coherently as a whole
\cite{SR1993,BBR99}.
If then the driving force or velocity decreases back
to a smaller value,
the film solidifies again
either in the solid state in the case of
spherical molecules,
or in an amorphous (glassy) phase for the case
of lubricant consisting of long-chain (organic) molecules
\cite{Robbins}.
Thus, if the top substrate is coupled through a spring
to a stage moving with a constant velocity,
then at the beginning the spring stretches and the force increases.
Eventually $f$ exceeds $f_s$, the top substrate
begins to slide and the lubricant melts.
Then the top substrate accelerates to catch up
with the stage and $f$ decreases,
so the substrates stick again.
This periodic melting--freezing process was used
by Thompson and Robbins
\cite{TR1990a}
for explanation of the stick--slip motion observed experimentally.
Forth,
when the film consists of flexible chain molecules,
then it may be trapped in a glassy state,
and the sliding occurs only
at the lubricant/substrate interfaces (the plug-like flow)
\cite{Robbins}.
This explains why
the effective viscosities of thin films may rise
more than five orders of magnitude above bulk values
(experimentally, the relaxation times are fractions of a second
for thin confined films and nanoseconds for a bulk).
Finally, the regime of smooth sliding
with atomic-scale velocities was studied in
\cite{SRC1996,BDP97,BBR99}.
It was shown that the kinetic frictional force
can be described by the introduction of
a phenomenological parameter which describes
an ``intrinsic'' damping within the lubricant
due to anharmonic coupling between its different modes.
Also, He and Robbins in a recent paper \cite{HR2000}
studied the friction at low velocities.
Their simulations lead to a frictional force which is
proportional to the load in accordance with
the Amontons law,
and rises logarithmically with velocity,
again in agreement with experiments.
The simulation results were explained as
thermally activated motion of the lubricant over
the periodic substrate potential.

However, all these studies
opened the question of the validity of results obtained
with simple models, in particular which results are of general validity and
which ones are model dependent, and to what extend the results
obtained on microscopic samples can be used to analyze actual
experiments performed on a macroscopic scale.
We think that these questions are important for the future of
molecular dynamics simulations of friction, and we want to address
some aspects of them in the present work.

A critical question is the energy flow out of the friction zone and
the way it is described in the models because it governs the value of
the friction coefficient.
Thermal equilibrium is rather easy to simulate correctly because it
does not depend on the value
of the external damping $\eta$ in Langevin equations
or on the method used to introduce the damping
(although the rate of approach to equilibrium
depends on $\eta$).
It is clear, however, that the non-equilibrium state of
a system driven by an external dc force
must depend on the method used to
take into account the energy exchange with the outside, for instance
on the type of  external damping which is assumed.
One may predict that the dependence is not
crucial if the rate of energy exchange between
different modes of the system,
which emerges due to nonlinearity of the equations of motion,
is (much) larger than energy transfered outside of the
system, described by the external damping $\eta$.
This case corresponds, for example, to a lubricant close
to the melting temperature studied in
\cite{HMR1999-MR2000b-MR2000a,TR1990a,SR1993,Robbins,SRC1996,HR2000},
especially for lubricants with complex 
(long--chain) molecules.
But for a monatomic lubricant or a lubricant with simple molecules
at low temperatures and low driving velocities,
when the lubricant film is in the solid state,
the simulation results may strongly depend
on the external damping.
In the present study we 
investigate this question by using a realistic energy
loss mechanism. 
Comparing with models previously used in friction simulation
(as discussed above, see also \cite{Robbins2} and references therein),
the main new feature of the present study is that
we use Langevin equations with 
an external damping coefficient $\eta(z,v)$
which depends on the distance $z$ of a lubricant atom
from the substrate and on its velocity $v$
with respect the substrate. 
Some significant differences with simpler modeling of the energy losses
are exhibited.

With this model, we pay a particular
attention to the motion at small velocity $v_{\rm top}$ of the
top substrate, looking for the 
smallest possible value of the velocity for which the motion stays
smooth.  We show that the microscopic transition from smooth sliding
to stick-slip motion takes place at a velocity which is {\it many orders of
magnitude higher than that observed in macroscopic experiments.}  
This is an important point in the connection between simulations 
and experiments, which
suggests that the macroscopic mechanism of the transition from
stick-slip motion to smooth sliding is completely different from the
microscopic one.

We also used our extended model to examine various aspects of
friction, which will be presented in the result and discussion sections.
The paper is organized as follows.
First, in Sec.\ \ref{model} we describe the model and
the algorithm used in the simulations, and we
introduce an external damping coefficient for Langevin equations,
based on microscopic considerations, which attempts to
describes the energy exchange between
the lubricant atoms and the substrates in a realistic way.
Then in Sec.\ \ref{results} we present the simulation results
and analyze them considering the excitation of vibrations in the system.
Finally, Sec.\ \ref{discussion} discusses the applicability
of microscopic simulation to macroscopic experiments.

%==================================================================
\section{Model}
\label{model}

We are interested in the regime of boundary lubrication,
when two sliding surfaces come in solid contact,
and in the contact region, the surfaces are separated by at most
a {\it few} layers of lubrication molecules.
We consider a model with a thin (one to five layers)
{\it solid\/} lubricant film between two solids.
There are two reasons for this choice.
{\it First},
there is always some lubricant 
(called ``third bodies'' by tribologists)  between the surfaces.
It may correspond either to a specially chosen lubricant, or to 
water or other material adsorbed from air,
a dust, grease, wear debris produced by sliding, {\it etc}.
In general the lubricant is not commensurate with the substrates,
and, even if they are commensurate, 
the lubricant and substrates are seldom perfectly aligned.
The external load squeezes the lubricant out from the contact area,
so a very thin lubricant film is often formed.
The last layer is however very difficult to remove \cite{P00,Plast}
so that there is at least one lubricant layer between the substrates.
Due to the load, the lubricant is always strongly compressed and
thus has to form a closely packed structure.
As a result, the confined thin film solidifies 
\cite{Robbins2,Robbins}
as was mentioned above.
{\it Second}, we are interested in low friction situations, 
and it is in the case of a solid thin lubricant film that we may expect
to obtain the minimal friction.
When the lubricant is liquid or amorphous, 
the kinetic friction is usually much larger 
\cite{Mate95,Gen,Gen2,SRC1996,Krim}.

Moreover, contrary to our previous studies 
\cite{BDP97,BBR99}, now the surface layers of the substrates are simulated
directly. The substrates consists of two parts. A rigid part forms the
boundaries of the model system, and a deformable substrate layer
is in contact with the lubricant, providing a much more realistic
description of the substrate than previously.

\begin{figure}
\epsfxsize=\hsize
\epsfbox{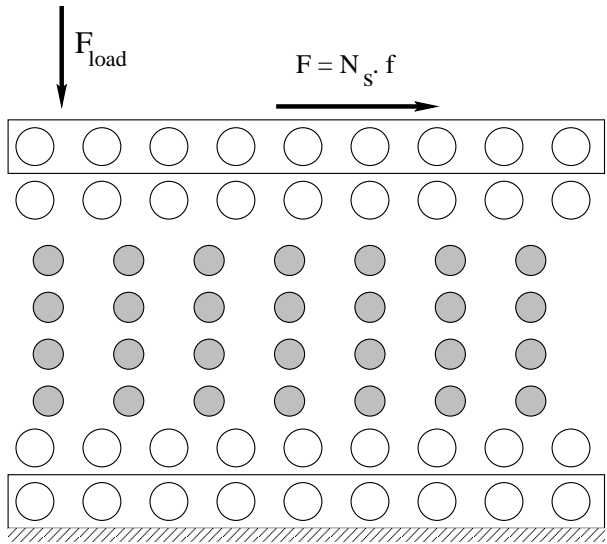}
\caption{
The model.}
\label{fig000}
\end{figure}

Our three-dimensional system comprises 
a few atomic-layer film
between two parallel rigid top and bottom substrate parts
as shown in Fig.\ \ref{fig000}.
Each rigid substrate part has $N_S=132$ atoms henceforth called s-atoms
organized into a $12 \! \times \! 11$ lattice
of square symmetry with the lattice constant $a_s=3$.
The atoms of the bottom rigid substrate part are fixed
while the top substrate part moves rigidly.
Between the rigid substrate parts we insert atoms of two
different kinds: $2 N_S$ s-atoms 
model the surfaces of the substrate, which are therefore
modeled more accurately than with rigid body, and $N_{al} N_l$
l-atoms (``lubricant'' atoms) form the lubricant film.
In the $x$ and $y$ directions we use periodic boundary conditions. 
The top substrate part and all the atoms may move in all three dimensions. 
To shorten the expressions we 
henceforth call top and bottom ``substrates'' the rigid parts of
the substrates. It should be kept in mind that these ``substrates''
are in fact only a part of the top and bottom solids that slide on
each other.
All atoms interact via a 6-12 Lennard-Jones (LJ) 
pairwise potential
\begin{equation}
V(r)=V_{\alpha \alpha^{\prime}} \left[
\left( \frac{r_{\alpha \alpha^{\prime}}}{r}
\right)^{12} -2
\left( \frac{r_{\alpha \alpha^{\prime}}}{r}
\right)^{6} \right],
\label{m0}
\end{equation}
but the parameters of the potential (\ref{m0}) are different for
different kinds of atoms. 
For the s-s interaction we took $V_{ss}=3$ and $r_{ss}=3$,
for the l-l interaction, $V_{ll}=1$ and $r_{ll}=4.14$
(``incommensurate'' case), and
for the s-l interaction, $V_{ls}=1/3$
(corresponding to a weak interaction between the substrate and the
lubricant) and 
$r_{ls}=\frac{1}{2} (r_{ss}+r_{ll})=3.57$,
respectively.
The cutting radius for the interaction
was chosen as $r^{*}=1.49\, r_{ll}=6.17$.
With these energy parameters,
the s-atoms stick to the top and bottom substrates, 
covering them with mono-layers, 
while the l-atoms 
fill the space between the substrates.
This simulates a lubricant between two solids,
and the surface (utmost) layer of each substrate
is treated explicitly.
The equilibrium configuration of the lubricant
corresponds to $N_l$ layers, each having $N_{al}=80$ atoms,
organized into a close-packed triangular lattice slightly
distorted by the substrate potentials.

For the masses we took $m_l=m_s=m_S=1$ which gives
a typical frequency for the system of 
$\omega_s = \left[
V_{ss}^{\prime \prime} (r_{ss})/m_s
\right]^{1/2} =4.9$,
and the corresponding characteristic period is
$\tau_s=2\pi/ \omega_s=1.28$.
A typical frequency of vibrations within the lubricant is
$\omega_{l}=6 \left( 2V_{ll} / m_{l} r_{ll}^2 \right)^{1/2}=2.05$.
To each atom of the top rigid substrate we apply a dc force % ${\bf F}$,
consisting of a driving force $f$ along the $x$ axis and
a loading force $f_{\text{load}}=-0.1$ 
along the $z$ direction.
These parameters have been used in most of the simulations,
except when otherwise specified.

\bigskip
{\it Units.\/}
Although we work with dimensionless quantities,
the numerical values of the model parameters 
have been chosen such that, 
if energy were measured in electron volts 
and distances in Angstr\"{o}ms, 
we would have realistic values for a typical solid. 
However, for a discussion of the applicability of simulations
to real physical systems,
it is useful to couple the ``natural units'' ($n.u.$) used in the simulation
with, e.g., the SI system of units.
The basic parameters which were unchanged in
all simulations, are the amplitude of interaction within the substrates
($V_{ss}=3$), which sets the energy parameter,
the substrate lattice constant
($a_{s}=3$) which sets the length scale,
and the mass of lubricant atoms
($m_l =1$) as the mass parameter.
A real system can be characterized by
the amplitude of interaction in the substrates
$\widetilde{V}_{ss}$ measured in eV,
by the substrate lattice constant
$\widetilde{a}_{s}$ measured in \AA,
and by the mass of lubricant atoms
$\widetilde{m}_l$ measured in proton masses $m_p$.
If we introduce the following coefficients,
$\nu_e = \widetilde{V}_{ss} / V_{ss} = \widetilde{V}_{ss} / 3$,
$\nu_r = \widetilde{a}_{s} / a_{s} = \widetilde{a}_{s} / 3$, and
$\nu_m = \widetilde{m}_l / (100 \, m_l) = \widetilde{m}_l / 100$,
then for a typical system we have
$\nu_e \sim \nu_r \sim \nu_m \sim 1$,
and it is easy to find these coefficients for a real physical system.

Now we can couple our natural units
with the SI system of units. Namely, we have
for the unit of length
$1 \; m = 10^{10} \, \nu_r^{-1} \; n.u.$,
for the unit of mass
$1 \; kg = 6 \cdot 10^{24} \, \nu_m^{-1} \; n.u.$,
for the unit of energy
$1 \; J = 6.25 \cdot 10^{18} \, \nu_e^{-1} \; n.u.$,
for the unit of force
$1 \; N = 6.25 \cdot 10^{8} \, (\nu_r / \nu_e) \; n.u.$,
for the unit of pressure
$1 \; N/m^2 = 6.25 \cdot 10^{-12} \, (\nu_r^3 / \nu_e) \; n.u.$,
for the unit of time % (1 J = 1 kg$\cdot$m$^2$/sec$^2$)
$1 \; s = 0.98 \cdot 10^{13} 
\left( {\nu_e}/{\nu_m \nu_r^2} \right)^{1/2} \; n.u.$,
and for the unit of velocity
$1 \; {m}/{s} = 1.02 \cdot 10^{-3} 
\left( {\nu_m}/{\nu_e} \right)^{1/2} \; n.u.$
In particular, the load force $f_{\rm load}=-0.1 \; n.u.$
used in the simulations, corresponds to the pressure
$P = -f_{\rm load}/a_s^2 =
  1.11 \cdot 10^{-2} \; n.u. =
  1.78 \cdot 10^9 \; N/m^2$.
To compare with experimentally used values, note that
a realistic pressure is $P \sim 10^7 \; N/m^2$, 
and the maximum pressure above which
the plastic deformation begins, is
$P \approx 2 \cdot 10^8 \; N/m^2$ for gold (a minimal value for metals),
$P \approx 10^9 \; N/m^2$ for steel, and
$P \approx 10^{11} \; N/m^2$ for diamond (the largest possible value).
As for velocities, a typical value when the transition
from a stick-slip motion to smooth sliding is observed
experimentally is
$v_c \sim 1 \; \mu m/s =  10^{-9} \; n.u.$
\cite{Mate95}.

%------------------------------------------------------------------
\bigskip
{\it Equations of motion.\/}
We use Langevin equations for all mobile atoms, i.e.\ atoms that do not
belong to a rigid substrate,
\begin{equation}
m_{\alpha} \ddot{r}_{i \alpha} =
f_{i \alpha}^{\rm (int)} +
\sum_{S=1}^{2}
f_{i \alpha, S},
\label{m1}
\end{equation}
where 
$r_{i \alpha} \equiv \left\{x_{i \alpha},y_{i \alpha},z_{i \alpha}\right\}$
is the coordinate of the $i$th atom
and $\alpha={\rm s}$ or $\alpha={\rm l}$
for ``substrate'' or ``lubricant'' atoms respectively.

The force $f^{\rm (int)}$ is due to interaction
between the mobile atoms in the system,
\begin{equation}
f_{i \alpha}^{\rm (int)} = 
-\frac{\partial}{\partial r_{i \alpha}}
\sum_{i^{\prime} \alpha^{\prime}}^{\rm all}
V_{\alpha^{\prime} \alpha}(r_{i^{\prime} \alpha^{\prime}} - r_{i \alpha}),
\label{m2}
\end{equation}
where the sum includes all ``mobile'' s- and
l-atoms except the given ($\alpha$th) one.

The last term in Eq.\ (\ref{m1}) describes the
interaction of a ``mobile'' s- or l-atom
with the bottom ($S=1$) and top ($S=2$) substrates.
The force $f_{i \alpha, S}$ itself
consists of three contributions as usual in Langevin equations,
\begin{equation}
f_{i \alpha, S} =
f_{i \alpha, S}^{\rm (int)} +
f_{i \alpha, S}^{\rm (fric)} +
f_{i \alpha, S}^{\rm (ran)}.
\label{m3}
\end{equation}
The first contribution
$f_{i \alpha, S}^{\rm (int)}$
comes from the potential interaction
of a given $i$th atom with
all ``immobile'' atoms of the $S$th
(bottom or top) rigid substrate,
\begin{equation}
f_{i \alpha, S}^{\rm (int)} = 
-\frac{\partial}{\partial r_{i \alpha}}
\sum_{i^{\prime}=1}^{N_S}
V_{s \alpha} (R_{i^{\prime} S} - r_{i \alpha}),
\label{m4}
\end{equation}
where the sum now includes all ``immobile'' 
s-atoms of the corresponding substrate
and $R_{i S}$ is the coordinate of the $i$th
atom of the $S$th rigid substrate.

The second and third contributions
in Eq.\ (\ref{m3}) describe the energy exchange
between mobile atoms and the rigid substrates. They
approximately take into account the
missing degrees of freedom of the substrates.
More precisely, the second contribution
$f_{i \alpha, S}^{\rm (fric)}$
describes a viscous damping when an atom
moves relative the corresponding substrate,
\begin{equation}
f_{i \alpha, S}^{\rm (fric)} =
-m_{\alpha} \eta (\ldots)
\left( \dot{r}_{i \alpha} - \dot{R_S} \right),
\label{m4b}
\end{equation}
where $\eta (\ldots)$ is the ``external'' damping coefficient
which depends on the velocity and distance
relative to the substrate,
\begin{eqnarray}
&\eta (\ldots) = \eta \left( z_{\rm rel}, v_{\rm rel} \right),
\nonumber \\
&z_{\rm rel} = (-1)^{(S-1)} (z_{i \alpha} - Z_S),
\;\;\;
v_{\rm rel} = \dot{r}_{i \alpha} - \dot{R_S},
\label{m4a}
\end{eqnarray}
$R_S \equiv \left\{ X_S,Y_S,Z_S \right\}$
is the center of mass coordinate of the $S$th
substrate (for the bottom substrate
we took $R_1 \equiv 0$).
Finally, the third contribution
$f_{i \alpha, S}^{\rm (ran)}$
in Eq.\ (\ref{m3})
describes the random (Gaussian) force acting
on the $i$th atom from the $S$th substrate.
The amplitude of this force is determined
by the substrate temperature $T$,
i.e.\ the corresponding correlation function is
\begin{eqnarray}
&\langle
f_{i \alpha, S}^{\rm (ran)} (t)
f_{i^{\prime} \alpha^{\prime}, S^{\prime}}^{\rm (ran)} (t^{\prime})
\rangle =
\nonumber \\
&2 \eta_R (\ldots)
m_{\alpha} k_B T
\delta_{i i^{\prime}}
\delta_{\alpha \alpha^{\prime}}
\delta_{S S^{\prime}}
\delta (t-t^{\prime}).
\label{m5}
\end{eqnarray}
The function $\eta_R(z_{\rm rel})$
in Eq.\ (\ref{m5})
coincides with the external damping coefficient
$\eta (z_{\rm rel})$
if the latter does not depend on the velocity
(but may depend on the coordinate).
Otherwise, if the external damping depends on $v_{\rm rel}$,
the two coefficients are coupled by the relationship
\cite{Gardiner,Braun-new}
\begin{eqnarray}
&\eta_R (z, v, T) = \int_0^{\infty} d\epsilon \;
e^{-\epsilon}
\eta (z, \widetilde{v} (\epsilon)),
\nonumber \\
&\widetilde{v}^2 (\epsilon) = v^2 +
\frac{2 k_B T}{m_{\alpha}}
\epsilon.
\label{m5a}
\end{eqnarray}

For the top rigid substrate we use
Newton equation of motion,
\begin{equation}
M_S \ddot{R}_{2} = N_S f_{\rm ext} +F_S,
\label{m10}
\end{equation}
where 
$M_S=N_S m_S$ is the mass of the rigid substrate,
$f_{\rm ext} = \{ f, 0, f_{\rm load} \}$
is the external force applied to it, and
$F_S=-\sum_{i\alpha}^{\rm all} f_{i\alpha, S=2}$
according to third Newton law
(conservation of the total momentum of the system).
As we checked numerically, this technique leads
to Gaussian distribution of velocities for all atoms
as well as for the top rigid substrate with a correct width
for a given temperature $T$.

%------------------------------------------------------------------
\bigskip
{\it External friction coefficient.\/}
Fortunately, the motion of a single atom or a sub-monolayer film
adsorbed on a crystal surface,
namely their vibration near the equilibrium position,
has been well studied experimentally and theoretically
\cite{vibrat,B1}.
This allows us to model the external damping
of lubricant atoms near the substrates
with a reasonable accuracy.
For the external damping coefficient
$\eta(z,v)$ which models the energy loss of an atom into the
substrates and enters into Eqs.\ (\ref{m4b}-\ref{m5}),
we take into account its dependence on the distance
$z$ from the corresponding substrate
and on the relative velocity $v$
according to
$\eta(z,v) = \eta_1(z) \eta_2(v)$.
First, we assume that the damping rate exponentially
decreases when an atom moves away from the substrate,
and saturates at some level when the atom approaches 
very close to the substrate,
\begin{equation}
\eta_1 (z) = 1-
{\rm tanh} [(z-z^{*})/z^{*}].
\label{m6}
\end{equation}
The characteristic distance $z^{*}$ is chosen as the height
of the pyramid with the base constructed of
four rigid substrate atoms with square length
$a_s=3$, and the vertex with the s-atom
at the distance $r_{ss}=3$ from the rigid-substrate atoms;
this leads to $z^{*}=2.12$.
Thus, for the atoms in the s-layer, where
$z \sim z^{*}$, we have $\eta_1 \sim 1$,
while for the atoms in the first 
(closest to the substrate) lubricant layer
we obtain $\eta_1 \sim 0.1$.

Second, to determine $\eta_2 (v)$
we used the results known for the damping of the vibrations
of an atom adsorbed on the crystal surface. According to the theory
\cite{B1,B2}, when an atom vibrates with a frequency
$\omega$ near its equilibrium position,
the oscillation decays due to creation of phonons
in the substrate with the rate
\begin{equation}
\eta_{\rm ph}(\omega)=\frac{\pi}{2}
\frac{m_{\alpha}}{m_S} \omega^2
\rho (\omega),
\label{m7}
\end{equation}
where the surface local density of phonon
states may be described approximately 
by the function \cite{B2}
\begin{equation}
\rho (\omega) = \frac{32}{\pi} \frac{\omega^2
(\omega_m^2 -\omega^2)^{3/2} }{\omega_m^6},
\label{m8}
\end{equation}
and $\omega_m$ is the maximum (Debye) phonon frequency
of the solid substrate.
The one-phonon damping mechanism works for frequencies
$\omega < \omega_m$. At higher frequencies the damping
is due to multi-phonon processes. Moreover,
in the case of metal or semiconductor substrates
there is an additional damping due to the creation of
electron-hole pairs in the substrate so that
the corresponding damping coefficient is typically
of order $\eta \sim 10^{-2} \omega_m$
(see \cite{B1,B2}).
Using these results, we took for the dependence
$\eta_2 (v)$ the following approximate expression,
\begin{equation}
\eta_2 (v) = \eta_{\rm min} +
\eta_{\rm ph} (2\pi v/a),
\label{m9}
\end{equation}
with $a=a_s$ for the motion along the substrate
and $a=z^{*}$ for the motion in the $z$ direction.
For the minimal contribution $\eta_{\rm min}$ we chose
$\eta_{\rm min} = 0.01 \, \omega_s =0.049$.
This value is compatible with the justifications given above,
and it leads to a thermalization of the system
in reasonable simulation times
$t < \! 3\cdot \! 10^3 \tau_s$.
The cutoff (Debye) frequency was found from
the phonon spectrum (the Fourier transform of velocities
of all atoms and the top rigid substrate 
calculated for $f=0$ at $T=0.025$) which gives $\omega_m=15$.
Thus, as a function of frequency
the external damping behaves as
\begin{equation}
\eta_2 (\omega) = \eta_{\rm min}
+16\, \omega_m \,
\left( \frac{\omega}{\omega_m} \right)^4
\left[1-
\left( \frac{\omega}{\omega_m} \right)^2
\right]^{3/2},
\label{eta2}
\end{equation}
so that
at small frequencies, $\omega \ll \omega_m$,
we have
$ \eta_2 (\omega) 
\approx \eta_{\rm min}+(16/\omega_m^3) \, \omega^4
\approx (4.9+0.47 \, \omega^4)
\!\cdot\! 10^{-2}$,
while at the frequency 
$\omega =  (4/7)^{1/2} \omega_m =11.34$
the external damping reaches its maximum value
$\eta_{2} \approx 22$.

%------------------------------------------------------------------

\bigskip
{\it Energy losses.\/}
When one part of a system moves with respect to another part with a
relative velocity $v$, then the rate at which work is done is equal
to $\varepsilon = v f$, where $f$ is the total force acting on the
former. This may be used to define the energy losses of a given atom as
\begin{equation}
\varepsilon_i = -\frac{1}{2} \sum_{i^{\prime}}
(v_i -v_{i^{\prime}}) f_{i i^{\prime}},
\label{m10a}
\end{equation}
where the sum is over {\it all\/} atoms of the system
including the s-atoms of the rigid substrates,
and $f_{i i^{\prime}}$ is the force acting on
the $i$th atoms from the $i^{\prime}$th one.
Note that this force has to include
the damping and random (Gaussian) contributions of the total
force when $i^{\prime}$ corresponds to the
s-atom of rigid substrates.
Then, taking a corresponding sum over atoms
and averaging over time,
one can find the energy losses in a given atomic
layer of the lubricant, or those in the substrates,
as well as separate contributions from
different degrees of freedom.

To illustrate this definition, let us consider
an ideal case, when the bottom substrate is immobile,
the top substrate moves with a velocity $v$
in the $x$ direction, and
the lubricant film with the velocity $\frac{1}{2} v$
while the force $F$ is applied to the top substrate. We have
$F_{\rm lub/top}=
-F_{\rm top/lub}=
 F_{\rm bot/lub}=
-F_{\rm lub/bot}=F$,
and we obtain
$\varepsilon_{\rm top} =
-\frac{1}{2} (-F) \left( v-\frac{1}{2}v \right) =
\frac{1}{4} Fv$,
$\varepsilon_{\rm lub} =
-\frac{1}{2} \left[ F \left( \frac{1}{2}v -v \right)
-F \frac{1}{2} v \right] =
\frac{1}{2} Fv$,
$\varepsilon_{\rm bot} =
-\frac{1}{2} F \left( 0-\frac{1}{2}v \right) =
\frac{1}{4} Fv$,
thus the total losses are equal to $\varepsilon =Fv$
which coincides with the work done by the external force as expected.

%------------------------------------------------------------------
\bigskip
{\it Curved substrates.\/}
In most simulations both top and bottom substrates were flat.
However, we have done also a few runs with curved substrates \cite{P1},
where the $z$ coordinate of the bottom rigid substrate
is defined by
\begin{eqnarray}
z = 
     - \frac{1}{2} h_x^{\rm (dn)} r_{sl}
\left( 1-\cos \frac{2 \pi x}{L_x} \right)
\nonumber \\
     - \frac{1}{2} h_y^{\rm (dn)} r_{sl}
\left( 1-\cos \frac{2 \pi y}{L_y} \right),
\end{eqnarray}
where $L_{x,y}$ is the size of the system
in the $x$ or $y$ direction
and $h_{x,y}$ are the corresponding curvature parameters.
Similarly 
the $z$ coordinate of the top rigid substrate
is defined by the expression
\begin{eqnarray}
z = Z_2
     + \frac{1}{2} h_x^{\rm (up)} r_{sl}
\left[ 1-\cos \frac{2 \pi (x-X_2)}{L_x} \right]
\nonumber \\
     + \frac{1}{2} h_y^{\rm (up)} r_{sl}
\left[ 1-\cos \frac{2 \pi (y-Y_2)}{L_y} \right],
\end{eqnarray}
where \{ $X_2$, $Y_2$, $Z_2$ \} 
are the center of mass
coordinates of the top rigid substrate.

\bigskip
%------------------------------------------------------------------
{\it Algorithm.\/}
In most of the simulations we used the constant-force algorithm 
\cite{BDP97,BBR99}. Namely,
at the beginning we equilibrate the system at a temperature $T$
(in this paper we present the results for
$T=0$ and $T=0.025$ which corresponds
to room temperature for energies measured in eV).
Then we adiabatically increase the shear force $f$
in steps $\Delta f$ (typically $\Delta f=10^{-3}$ or smaller).
Each increase of $f$ is accomplished progressively in
$i_t$ (typically $i_t=200$ or more) substeps each of duration 
$\tau_s$.
When $f$ has reached its new value, 
we wait during a time $i_t \tau_s$ in order to allow 
the system to reach a stationary state, 
and then measure the atomic coordinates and velocities 
during the next time interval $i_t \tau_s$.
Thus, the force is changing with the average rate
${\cal R}/3\tau_s$, where
${\cal R}=\Delta f / i_t$.
In some simulations we also used the
constant-velocity algorithm 
(which is typically used in tribology simulations,
see \cite{Jeam}),
when a constraint is used to keep constant the
top substrate velocity.
We found, however, that the const-$v$ algorithm leads
to a much smaller accuracy
of the results than the const-$f$ one,
probably because the friction force in this case
is not a self-averaging value.
Thus, here we present the results of the const-$f$ runs only.

\bigskip
Most of the simulations have been done for a rather small system.
The issue of system size raises the usual question in computer
simulations of whether it is better to take a small system and
study it carefully or to take a larger system and do a less
comprehensive study over shorter times.
Our simulations show that the processes under investigation
are characterized by long relaxation times.
The approach with adiabatic changing of the driving force 
necessitates going through a whole cycle of $f$ changes.
For a much larger system it is not possible to reach equilibrium
in a reasonable time for this type of study.
However, we also made few runs for a system four times larger
and found that the results did not change.
Besides, in physical systems microscopic contacts are of a size similar to
that of our system, for example $10 \times 10$ atoms. 
Even in  specially  prepared
surfaces, with experiments performed in high vacuum conditions,  ideal
surfaces usually do not extend beyond about 30 lattice constants.
Of course the results deduced from such simulations of ideal surfaces
should not be used without extreme caution to analyze the results of
macroscopic experiments. We shall come back to this point in the
discussion because the connection with macroscopic experiments is a
difficult point for all studies of microscopic friction, whether they
are experimental (with atomic force microscope) or numerical.

%==================================================================
\bigskip
\section{Results}
\label{results}

\begin{figure}
\epsfxsize=\hsize
\epsfbox{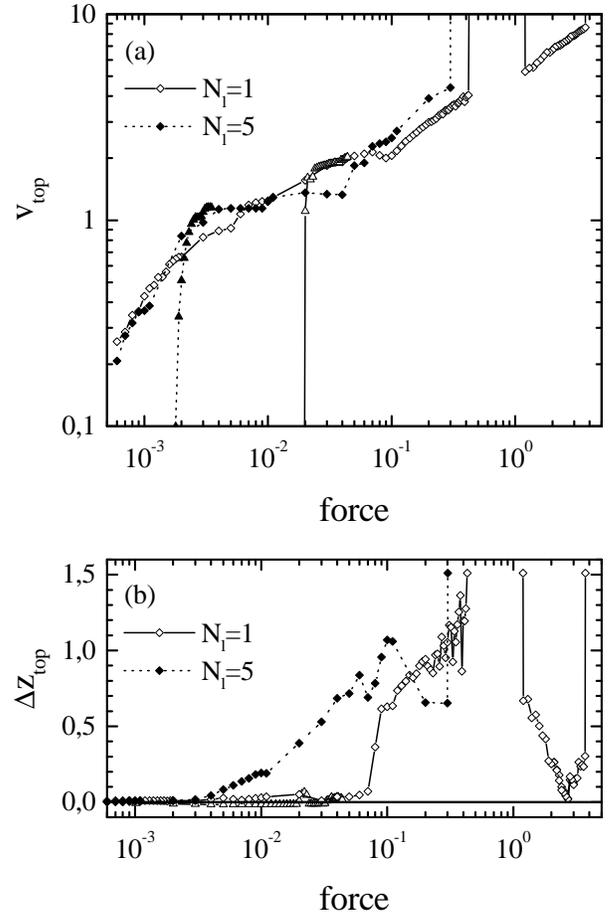}
\caption{
(a) The velocity of the top substrate and
(b) the change of the lubricant thickness
for the one-layer
(solid curves and open diamonds)
and five-layer
(dotted curves and solid diamonds)
lubricant films.}
\label{fig010}
\end{figure}

Before going into the details of simulation results,
let us discuss some general features.
Typical dependences of the velocity of the top substrate
in the $x$ direction, $v_{\rm top}(f)$,
and the change of its $z$ coordinate, $\Delta z_{\rm top}(f)$,
as functions of the driving force $f$
for the one- and five-layer lubricant films are shown in
Fig.\ \ref{fig010}.
Each plotted point is the average over $i_t=200$ data points 
recorded at times separated by $\Delta t = \tau_s$
and corresponds to the steady-state motion for a given $f$.

Let us analyze first the behavior of the five-layer film.
The motion starts when the force exceeds
a static friction force $f_s  \approx 0.0017 - 0.0018$.
One can see a plateau on the $v_{\rm top}(f)$ dependence
for the force interval $2\!\cdot\!10^{-3} <f< 0.03$,
then at larger forces $v_{\rm top}$ approximately linearly increases
with $f$, and finally at $f \geq f_f \approx 0.30 - 0.31$
the ``regular'' motion is destroyed.
The velocity of the top substrate for these force intervals lies
within the interval $1<v_{\rm top}<4.5$,
so that the washboard frequency, defined as
$\omega_{\rm wash} = \pi \langle v_{\rm top} \rangle /a$,
lies just inside the phonon spectrum of the lubricant film
as will be discussed in details later.
In this force range, the width $z$ of the film increases 
as shown on Fig.\ \ref{fig010}b.
The system exhibits hysteresis:
if one starts with a force $f < f_f$
and then decreases it adiabatically slowly,
the system comes back to the $T=0$ ground state
for a force $f_b \approx 2\!\cdot\!10^{-5}$,
which is much lower than $f_s$.
The velocity of the top substrate at the backward transition
drops down abruptly from a value $v_b \approx 0.03$ to zero.
Moreover, if we start from the sliding state
and then remove the driving force at all,
the system again comes to the $T=0$ ground state.

\medskip
The behavior of thinner lubricant films is very similar.
For example, the one-layer film starts to move at
$f_s \approx 0.019 - 0.02$. The forward transition
now takes place at $f_f \approx 0.43$ 
which is slightly larger than that for the five-layer film,
although the velocity at which the film starts to melt
is approximately the same, $v_f \approx 4$.
The backward transition takes place at
$f_b \approx 1.5\!\cdot\!10^{-4}$ when 
$v_b \approx 0.085$.

\medskip

\begin{figure}
\epsfxsize=\hsize
\epsfbox{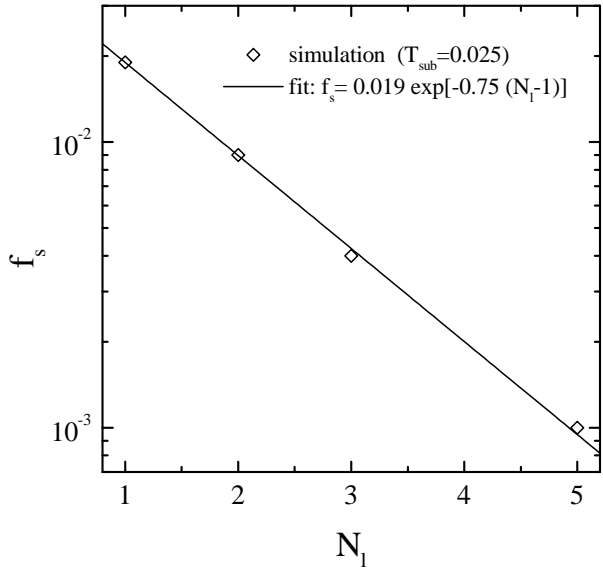}
\caption{
The static friction force $f_s$ as function
of the number of lubricant layers
$N_l$ for $T=0.025$.}
\label{fig020}
\end{figure}

Although the sliding regime is approximately the same
for different lubricant widths,
the static frictional force needed to initiate the sliding
decreases approximately exponentially as the width increases
as shown in Fig.\ \ref{fig020}.
The dependence $f_s (N_l)$ may be explained
qualitatively in the following way.
To initiate the sliding the lubricant atoms must
overcome barriers created by the substrate atoms.
But the height of the barriers strongly (exponentially)
depends on the distance $z$ of the lubricant atoms
from the substrate.
It decreases when the lubricant atoms are 
moved away from the substrate.
In the one-layer case
the confined lubricant has to adjust 
to both substrates simultaneously
and thus has no freedom to move
in the $z$ direction.
On the other hand, when the lubricant is thicker,
its utmost layers can move in the $z$ direction
due to elasticity of the film,
and thus can allow the atomic positions to find
a minimum-energy saddle configuration.
This is confirmed by the analysis of the system configurations 
just before $f_s$. They show 
that the shifts of the atoms in the $z$ direction
from their $f=0$ equilibrium positions
in the five-layer film are much larger than those in the
three-layer film which in turn are larger than in the
two-layer film.

\medskip
The dependence of the static force $f_s$ 
on the temperature $T$ is only significant for $N_l>1$
when the lubricant atoms have enough freedom to move
in the $z$ direction. For example, for the five-layer film
we found
$f_s \approx 0.0017$ at $T=0$ but
$f_s \approx 0.001$ at $T=0.025$.

\medskip

\begin{figure}
\epsfxsize=\hsize
\epsfbox{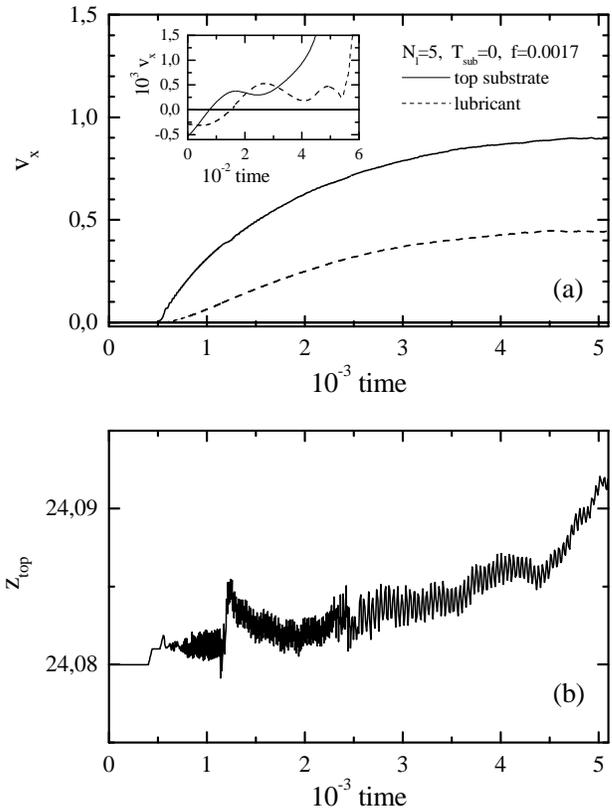}
\caption{
Beginning of the sliding of the five-layer film.
(a) The $x$-velocity of the top substrate
(solid curve) and the lubricant (dashed curve).
Inset: the same enlarged for shorter times.
(b) $z_{\rm top}$ versus time.}
\label{fig030}
\end{figure}

The initial stage of the sliding is illustrated on
Fig.\ \ref{fig030}, where we plot
$v_x(t)$ for the top substrate and the lubricant
and $z_{\rm top}(t)$ for the five-layer film.
One can see that at the beginning,
the top substrate
(the rigid substrate with one attached s-layer)
starts to move first.
Its velocity increases according to a law
$v_{\rm top}(t) \approx \left[f_s / (m_s+m_S) \eta_{\rm eff} (t) 
\right] \left[ 1-e^{-\eta_{\rm eff} (t)
(t-t_s)} \right]$,
where the effective friction coefficient is defined as
$\eta_{\rm eff}=f/(m_s+m_S)v_{\rm top}$,
and achieves a plateau in the steady state.
Soon after, however, the lubricant is also involved into
the motion and reaches the velocity
$v_{\rm lub} = \frac{1}{2} v_{\rm top}(f_s)$
in the steady state.
Simultaneously the width of the lubricant slightly increases.
We observed that for our system sizes,
the lubricant begins to move
as a whole. All its atoms begin to move almost simultaneously.
However, for larger sizes of the contact one could expect
that the sliding may begin with the creation of ``moving islands''
\cite{Joanna}.

\medskip
An interesting result is that the one-layer film
can be in the sliding steady state for much larger 
dc forces than  thicker films.
The solid sliding exists for forces
within the interval $1.1<f<3.7$ 
when the velocity is $5<v_{\rm top}<9$.
This steady state cannot be reached with by adiabatically increasing
the force because the lubricant melts earlier.
But it can be obtained with sharp increase of the force,
e.g.\ if one takes the $f=0.2$ steady state
and applies the  force $f=2$.
This high-speed steady state corresponds to the ``flying''
regime predicted in \cite{BBR99}
and will be described in more detail below in 
Sec.\ \ref{discussion}.

%------------------------------------------------------------------
\subsection{Solid sliding}

All steady states described above
correspond to a solid sliding regime,
in which the lubricant moves as a whole with a velocity
equal to half of the velocity of top substrate.
The distributions of velocities for all forces
can be well approximated by Gaussian curves if
we use different ``temperatures''
for the lubricant and the s-atomic substrate layers
as well as for different degrees of freedom.

\begin{figure}
\epsfxsize=\hsize
\epsfbox{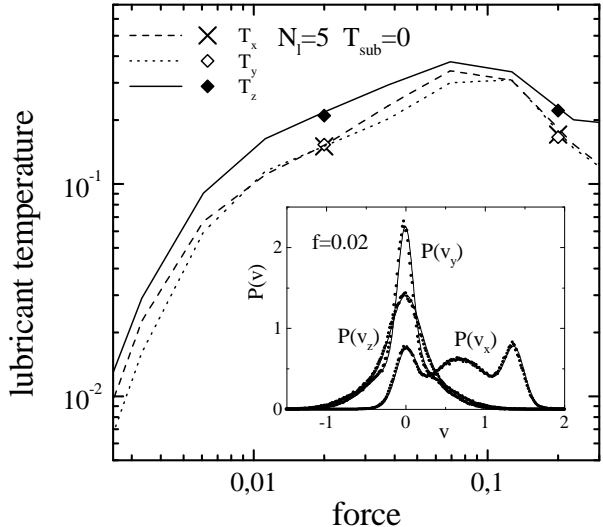}
\caption{
Lubricant temperatures versus force
for a five-layer film.
Inset: velocity histograms (dots)
and the corresponding Gaussian distributions
(solid curves) for $f=0.02$.}
\label{fig040}
\end{figure}

\begin{figure}
\epsfxsize=\hsize
\epsfbox{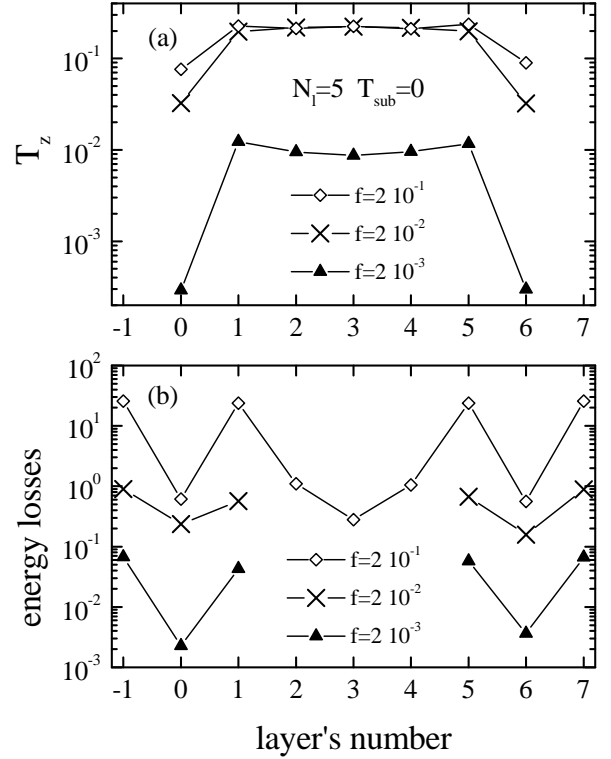}
\caption{
Distribution of (a) the temperature $T_z$
and (b) the energy losses across the five-layer lubricant
for different dc forces $f$ at $T_{\rm sub}=0$.
The numbers -1 and 7 correspond to the rigid substrates,
the numbers 0 and 6 correspond to the s-atomic substrate layers, and
the numbers 1 to 5 correspond to the lubricant layers.}
\label{fig050}
\end{figure}

For example, for the five-layer film we show 
in the inset of Fig.\ \ref{fig040}
the velocity histograms for the $f=0.02$ case
by dotted curves
and the corresponding Gaussians by solid curves.
Thus, we can define the ``temperature'' for a given
degree of freedom ($\alpha=x,y,z$) as well as for
a given layer $l$ of the lubricant or the substrate as
$T_{\alpha}=m \langle \left( 
v_{\alpha} - \langle v_{\alpha} \rangle 
\right)^2 \rangle$,
where $\langle \ldots \rangle$ designates
the averaging over time and over all atoms in the given layer,
and then use the values $T_{\alpha}(l)$
(let $l=0$ corresponds to the s-atomic layers
of the bottom substrate,
$l=1$ corresponds to the first layer of the lubricant
film, etc.)
as reliable characteristics of the driven system.
The distribution of temperatures across the system
is shown in Fig.\ \ref{fig050}a.

First, note that comparing with the lubricant temperatures,
the temperatures of s-layers of the substrates are very small
(for the $T=0$ simulation).
Second, at low forces/velocities
the temperature is not uniformly distributed over the lubricant film,
the boundary layers which are in moving contact with
the substrates, have a higher temperature than
those in a middle of the lubricant.
But at large forces/velocities,
$f \sim 0.02 - 0.2$ when $v_{\rm top} \sim 1 - 4$,
the lubricant temperature is approximately uniform
across the lubricant.
This indicates that anharmonicity effects,
which are responsible for energy exchange between
different layers within the lubricant,
become large enough at these forces.
Third, the lubricant temperature increases with growing $f$ 
(see Fig.\ \ref{fig040})
until it finally melts at $f=f_f$.
Forth, for all studied cases we found that
$T_z \gg T_x \agt T_y$.
This indicates that the driven system
is strongly out of equilibrium.

\medskip
The calculation of energy losses $E_{\rm loss}$ for different
dc forces shows that total losses are close 
to the expected values.
For example, for $f=0.002$ we obtained
$E_{\rm loss} \approx 0.24$, while $f v_{\rm top} \approx 0.26$;
for $f=0.02$, $E_{\rm loss} \approx 3.2$ ($f v_{\rm top} \approx 3.6$);
and for $f=0.2$, $E_{\rm loss} \approx 102$
($f v_{\rm top} \approx 103$) correspondingly.
The energy is lost mainly due to the motion of atoms along the
direction $x$ of the driving,
the $y$ and $z$ components of the losses are negligible.
The distribution of energy losses
in the normal direction of the system is
shown in Fig.\ \ref{fig050}b.
One can see that the energy is lost mainly
within the rigid substrates and in the utmost
lubricant layers (i.e.\ in the layers which are
in moving contact with the substrates).

\medskip

\begin{figure}
\epsfxsize=\hsize
\epsfbox{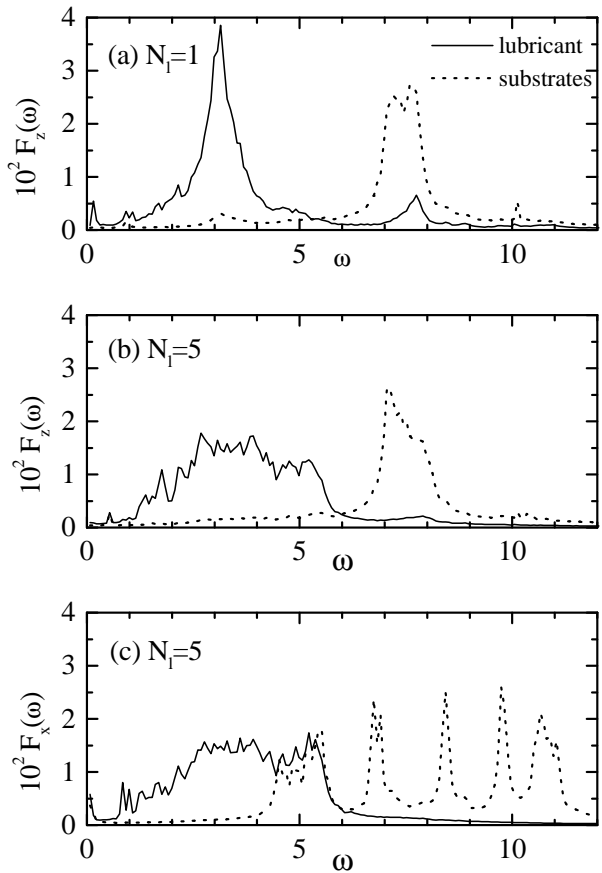}
\caption{
Phonon spectra of the lubricant (solid curves)
and the substrates (dotted curves)
for one- and five-layer films.}
\label{fig060}
\end{figure}

To analyze in details the energy losses for the solid-sliding regime,
we have to know the phonon spectrum of the system.
To find it, we saved the coordinates of all atoms
at $f=0$ and temperature $T=0.025$, and then calculated the Fourier
transform 
\begin{equation}
F_{\alpha} (\omega) =
\sum_{j}
\biggr| \int dt \, v_{j \alpha}(t) e^{i \omega t}
\biggr|,
\label{FFT}
\end{equation}
where the sum may be over all atoms
or only over a selected subset, such as the lubricant atoms.
The spectra calculated in this way
for different degrees of freedom ($\alpha=x,y,z$)
separately for the substrates and the lubricant
are shown in Fig.\ \ref{fig060} for $N_l=1$ and 5.
One can see that the lubricant spectrum occupies
the frequencies $1 \alt \omega \alt 6$, while the substrate modes
lie mainly within the interval $ 6< \omega < \omega_m$.
The $x$ and $y$ spectra are broader than the $z$ one.
For the one-layer lubricant, 
the spectrum $F_z(\omega)$ for the lubricant has
a maximum at $\omega \sim 3$ which is quite sharp.

\begin{figure}
\epsfxsize=\hsize
\epsfbox{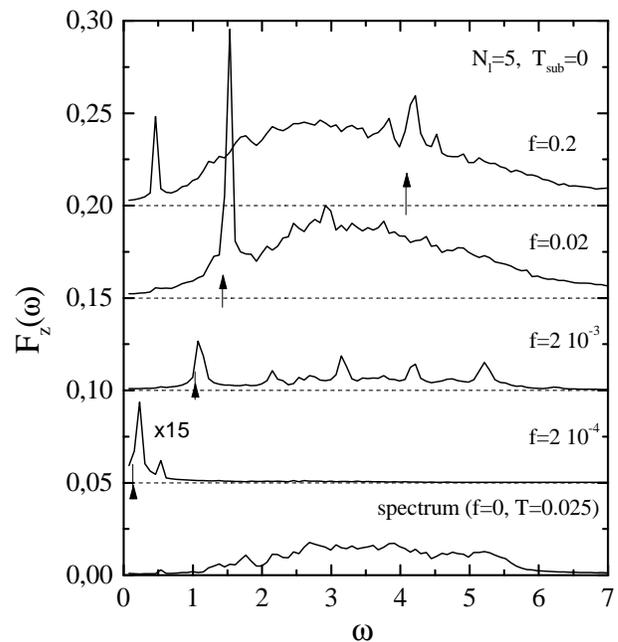}
\caption{
Change of the $z$-spectrum (\ref{FFT}) of five-layer
lubricant with increasing driving force.
The arrows indicate the position of the washboard frequency.
The curves for $f \neq 0$ are artificially shifted upstair.}
\label{fig070}
\end{figure}

\medskip
Calculating in the same manner the spectrum
in the solid-sliding regime,
we observed the
excitation of $z$-oscillations of the lubricant
with the frequency
$\omega = \omega_{\rm wash}$
as indicated by arrows in Fig.\ \ref{fig070}.
We also found that
(a) the intrinsic energy losses in the lubricant
become important only when
$\omega_{\rm wash}$ is higher than the lower
boundary of the lubricant phonon spectrum,
$\omega_{\rm wash} \agt 1$;
(b) at $f=2\!\cdot\!10^{-3}$ % we observe 
one can see
many higher-frequency harmonics of $\omega_{\rm wash}$, 
i.e.\ the vibration of lubricant
is highly anharmonic; when the force increases even more, the energy
exchange between the modes of the lubricant distributes the
energy among all the lubricant modes;
(c) at $f = 0.02$ and $f = 0.2$ 
the lubricant is strongly heated,
the energy of the translational motion is distributed over all three
($x$, $y$ and $z$) degrees of freedom.

\bigskip
The behavior of the thin $N_l=1$ film is similar.
The temperatures of the lubricant begin to increase
at $f>4\!\cdot\!10^{-3}$
when the velocity is $v_{\rm top}>1$ so that the washboard
frequency penetrates into the lubricant phonon spectrum.
Then at $f \agt 0.1$ the excitation of lubricant
vibrations strongly increases, especially for the
$z$ coordinate, so that the width of the film begins to increase
(see Fig.\ \ref{fig010}b). The temperature
reaches the value $T_z \sim 0.4$,
and soon the lubricant melts.
On the other hand, for the high-force sliding state
(the ``flying'' regime)
the perturbation of the film is much smaller,
$\Delta z < 0.5$ and $T_z \sim 0.1$.
At small forces/velocities the energy losses are mainly
in the rigid substrates.
At $f \sim 0.02 - 0.2$ the losses are approximately
equally distributed between the two rigid substrates
and the lubricant.
Finally, in the high-force ``flying'' regime the main
losses are in the lubricant film.

%------------------------------------------------------------------
\subsection{Effective friction coefficient}

In a steady-sliding regime, 
where the top rigid substrate and
one s-layer attached to it slide over the lubricant film,
it is convenient to introduce 
the effective friction coefficient as
\begin{equation}
\eta_{\rm eff}(v_{\rm top}) = \frac{f}{(m_S+m_s) v_{\rm top}(f)}.
\label{m11}
\end{equation}
Note that a change of the velocity of the top substrate
can be expected to take a time of the order
$\tau_{\rm relax}=\eta_{\rm eff}^{-1}$.
Using the results of simulations, one can see that
at small forces
the effective friction can be smaller than
$\eta_{\rm eff} < 10^{-3}$,
so the relaxation times are larger than
$\tau_{\rm relax} > 10^3$.

\medskip
Energy loss can only occur through the externally imposed damping.
One part comes from the motion of the lubricant as a whole.
This is the losses that one would get if the lubricant were sliding
as a perfect solid relatively the rigid substrates.
The corresponding contribution to the effective friction coefficient
is due to the external damping 
[with the coefficient $\eta(z,v)$
introduced into the equations of motion from the beginning]
of the uniform $x$-motion of lubricant atoms relative the substrates
with the velocity $\frac{1}{2} v_{\rm top}$,
and yields a ``universal'' dependence $v_{\rm top}^{\rm (uni)}(f)$
which does not depend neither on the lubricant width
nor on the masses $m_s$ and $m_S$.
In what follows we call this contribution as the
``perfect-sliding'' one.
A second part comes from the internal motions inside the lubricant
and the surface layers of the substrates. 
The driving excites atomic vibrations 
(mainly in the $z$ direction as mentioned above)
which then are distributed over other ($x$ and $y$) degrees of freedom
and finally are damped again
due to the external damping $\eta(z,v)$.
This contribution will be called below as the ``internal-losses'' one.

\medskip
In the solid-sliding regime,
when the top rigid substrate with one attached
s-layer moves as a whole with a velocity
$\langle v_{\rm top} \rangle$,
the bottom rigid substrate with one attached
s-layer does not move at all,
and the lubricant film moves as a whole with the velocity 
$v_{\rm lub} = \frac{1}{2} \langle v_{\rm top} \rangle$,
the washboard frequency is equal to
\begin{equation}
\omega_{\rm wash}
=2\pi v_{\rm lub}/a
=\pi \langle v_{\rm top} \rangle /a,
\label{wash}
\end{equation}
where one should take $a=a_s=3$.
Let us write the balance of forces
for the top substrate with one attached s-layer as
\begin{equation}
F \equiv N_S f
= N_{al} m_l \eta^*(v_{\rm top}) \, v_{\rm lub},
\label{bal}
\end{equation}
where we have introduced
the total viscous damping coefficient $\eta^*(v_{\rm top})$
for an atom in the utmost lubricant layer.
Using the definition (\ref{m11})
of the effective friction coefficient,
we obtain the following relationship between
the coefficients $\eta_{\rm eff}(v_{\rm top})$
and $\eta^*(v_{\rm top})$,
\begin{equation}
\eta_{\rm eff}(v_{\rm top}) = \frac{1}{2}
\, \frac{m_l}{m_S+m_s} \, \frac{N_{al}}{N_S} \, \eta^*(v_{\rm top}).
\label{eff}
\end{equation}
In a general case the total damping $\eta^*(v_{\rm top})$
defined by Eq.\ (\ref{bal}),
has to consist of two contributions,
the ``perfect-sliding'' contribution
$\eta_{\rm ext}(v_{\rm top})$ and
the intrinsic losses $\eta_{\rm int}(v_{\rm top})$.
In the two following subsection we evaluate successively these
two components of the losses.

%------------------------------------------------------------------
\subsubsection{``Perfect-sliding'' contribution to the energy loss}
\label{universal}

In the perfect-sliding approximation
the atoms in the utmost lubricant layers
feel only the external damping
$\eta_{\rm ext} (v_{\rm top}) \approx \eta_1 (z_{\rm lubr})
\, \eta_2 (\omega_{\rm wash})$
due to energy exchange with the rigid substrates.
Substituting the washboard frequency (\ref{wash})
into Eq.\ (\ref{eta2}), we obtain
\begin{eqnarray}
&\eta_{\rm ext} (v_{\rm top}) \approx 
\left[ 1-
{\rm tanh} \left( \frac{z_{\rm lubr}-z^{*}}{z^{*}} \right)
\right]
\nonumber \\
&\left[
\eta_{\rm min}+
\frac{16 \pi^4}{a_s^4 \omega_m^3} \, v_{\rm top}^4 \,
\left( 1-
\frac{\pi^2}{a_s^2 \omega_m^2} \, v_{\rm top}^2
\right)^{3/2}
\right].
\label{etaext}
\end{eqnarray}
Assuming that all the damping within the lubricant
is due to the external one, 
$\eta^*(v_{\rm top}) =\eta_{\rm ext}(v_{\rm top})$,
we obtain the ``universal'' (``perfect-sliding'') dependence
\begin{equation}
v_{\rm top}^{\rm (uni)} (f) = \frac{2 N_S}{N_{al}}
\frac{f}{m_l \eta_{\rm ext}(v_{\rm top})},
\label{etaNum}
\end{equation}
which does not depend neither on the number of lubricant layers
not on the substrate masses $m_s$ and $m_S$,
because it corresponds to the steady state
(while a delay of response of $v_{\rm top}$ when $f$ varies non-adiabatically,
has to depend on the masses).

\begin{figure}
\epsfxsize=\hsize
\epsfbox{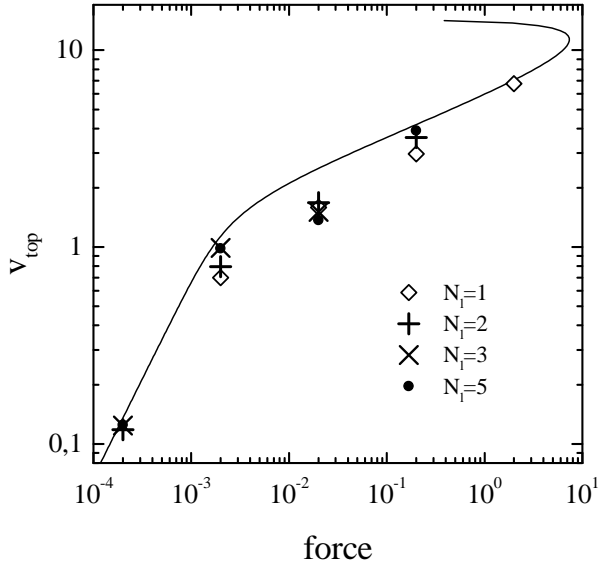}
\caption{
The ``perfect-sliding'' dependence (\ref{etaNum})
and the $T=0$ simulation results.}
\label{fig080}
\end{figure}

In particular, for the parameters used in simulation
($m_l=m_s=m_S=1$, $N_S=132$ and $N_{al}=80$), we obtain
$\eta_{\rm eff} = 0.15 \, \eta^*$.
Then, using in Eq.\ (\ref{etaext}) the values
$z^*=2.12$,
$\omega_m=15$,
$a_s=3$,
and taking from the simulation
$z_{\rm lubr} \approx 5.25$ for the $T=0$ ground state configuration,
we obtain
$\eta_{\rm ext} (v) \approx 
\left[ 4.9+0.57 \, v^4
\, \left( 1-0.49 \!\cdot\! 10^{-2} v^2 
\right)^{3/2} \right]
\!\cdot\! 10^{-3}$,
so that the external phonon damping
exceeds the minimal one (which models
the electron-hole damping)
at $v_{\rm top} \agt 1.7$
and reaches its maximum value
$\eta_{\rm ext} \approx 2.16$
at $v_{\rm top} \approx 10.8$.
If all the damping within the lubricant
were due to the external one, 
the effective friction coefficient would be equal to
\begin{eqnarray}
&\eta_{\rm eff}^{\rm (uni)} (v_{\rm top}) \approx
0.728 \!\cdot\! 10^{-3} \left[ 1+
\right.
\nonumber \\
&\left.
0.116 \, v_{\rm top}^4
\, \left( 1-0.49 \!\cdot\! 10^{-2} v_{\rm top}^2 
\right)^{3/2} \right]
\label{etaNum2}
\end{eqnarray}
with the maximum
$\eta_{\rm eff} \approx 0.32$.
The ``perfect-sliding'' dependence (\ref{etaNum}) 
is presented in Fig.\ \ref{fig080}.
One can see that it agrees rather well with
the simulation data at small ($f<10^{-3}$)
as well as at high ($f>1$) forces, when
the washboard frequency lies outside the lubricant phonon spectrum and, thus,
the internal motions of the lubricant are not excited.

\medskip
An important result of this analysis is that
the lubricant cannot support a force larger than
$f_{\rm max} \approx 7$ when $v_{\rm top} \approx 10.8$.
At larger forces/velocities the system cannot
dissipate the energy injected into it due to driving,
and the solid-sliding regime must be destroyed.

\medskip
However, in the simulation we found that
$v_{\rm top}(f) < v_{\rm top}^{\rm (uni)}(f)$ 
and $f_f < f_{\rm max}$ in all cases.
To describe simulation results more accurately,
the following three factors have to be taken into account:
({\it i\/}) at velocities $v_{\rm top} > 1 - 2$ the lubricant width increases
(due to anharmonicity of the LJ potential),
({\it ii\/}) there are additional intrinsic
energy losses within the lubricant
as well as in the s-layers, and
({\it iii\/}) due to these intrinsic energy losses,
the lubricant is heated during the sliding.
The first factor increase of $v_{\rm top}$ while
the second and third ones tend on the contrary to
decrease it so that they must be studied in details before a
conclusion can be drawn.

%------------------------------------------------------------------
\subsubsection{Excitation of lubricant vibrations}

When the lubricant film moves over the bottom substrate
with the velocity $v_{\rm lub}$
dragged by the top substrate with the dc force $F=N_S f$,
the dragging force per lubricant atom
in the utmost lubricant layer is equal to
$f_{\rm lub}=(N_{S}/N_{al}) f$.
The total energy injected into the lubricant
by the external dc force
per time unit per one lubricant atom,
is equal to
\begin{equation}
\varepsilon_{\rm tot}
=f_{\rm lub} v_{\rm lub} =m_l \eta^* v_{\rm lub}^2,
\label{res1}
\end{equation}
where we used the definition (\ref{bal})
of the coefficient $\eta^*$.
This energy must be absorbed in the system.
One channel of energy dissipation,
the energy exchange with the rigid substrates,
was already considered in the previous subsection,
and it yields
\begin{equation}
\varepsilon_{\rm ext}
=m_l \eta_{\rm ext} v_{\rm lub}^2.
\label{res2}
\end{equation}
An extra energy losses,
$\varepsilon_{\rm int} =
\varepsilon_{\rm tot} -
\varepsilon_{\rm ext}$,
must be attributed to intrinsic losses
within the system.

\medskip
To study the intrinsic losses,
we will use the technique similar to that described in
\cite{Sokol1,SRC1996,BDP97}. Let us
consider the internal vibrations of the
system as a damped oscillator
$\xi(t)$ of mass $m$, internal frequency $\omega_0$
and damping coefficient $\eta_0$, where
$\eta_0$ corresponds to the so-called
full width at half maximum (FWHM)
of the spectrum
(a generalization to many oscillators is trivial).
If the oscillator is excited by an external force
oscillating with the frequency $\omega$
and the amplitude $f_0$,
\begin{equation}
m \ddot{\xi} + m \eta_0 \dot{\xi} +
m \omega_0^2 \xi = f(t) = {\rm Re}
\left( f_0 e^{i \omega t} \right),
\label{res3}
\end{equation}
the steady-state motion of the oscillator corresponds
to its vibration with the frequency 
of the external oscillating force,
\begin{equation}
\xi(t) = \frac{f_0}{m} \, {\rm Re}
\left( \frac{e^{i \omega t}}
{\omega_0^2 -\omega^2 + i \omega \eta_0} \right).
\label{res4}
\end{equation}
The rate of energy losses
(the energy absorbed by the oscillator per one time unit) 
is equal to
\begin{equation}
\varepsilon_{\rm osc} (\omega; \; \omega_0, \eta_0, f_0) =
\frac{1}{\tau} \int_0^{\tau} dt \,
f(t) \dot{\xi}(t) =
\frac{1}{2} \, \omega \, \frac{f_0^2}{m} \, \rho(\omega),
\label{res5}
\end{equation}
where
$\tau =2\pi /\omega$ and
\begin{equation}
\rho(\omega) = {\rm Im} \left( \frac{1}
{\omega_0^2 -\omega^2 + i \omega \eta_0} \right) =
\frac{\omega \eta_0}
{(\omega^2 -\omega_0^2)^2 + (\omega \eta_0)^2}.
\label{res6}
\end{equation}

In the solid-sliding regime the oscillating force
is due to motion of the lubricant film
over the corrugated substrates and thus is
characterized by the frequency $\omega=\omega_{\rm wash}$.
Then, assuming that the system can be described as
$k$ different oscillators and
using Eqs.\ (\ref{res1}--\ref{res6}), we finally obtain
$\eta^*(v_{\rm top})=\eta_{\rm ext}(v_{\rm top})+
\eta_{\rm int}(v_{\rm top})$,
where
\begin{equation}
\eta_{\rm int}(v_{\rm top}) =
\frac{4}{m_l v_{\rm top}^2}
%m_l^{-1} (v_{\rm top}/2)^{-2}
\sum_{i=1}^{k}
\varepsilon_{\rm osc} (\omega_{\rm wash}; \;
\omega_0^{(i)}, \eta_0^{(i)}, f_0^{(i)}).
\label{res7}
\end{equation}

\begin{figure}
\epsfxsize=\hsize
\epsfbox{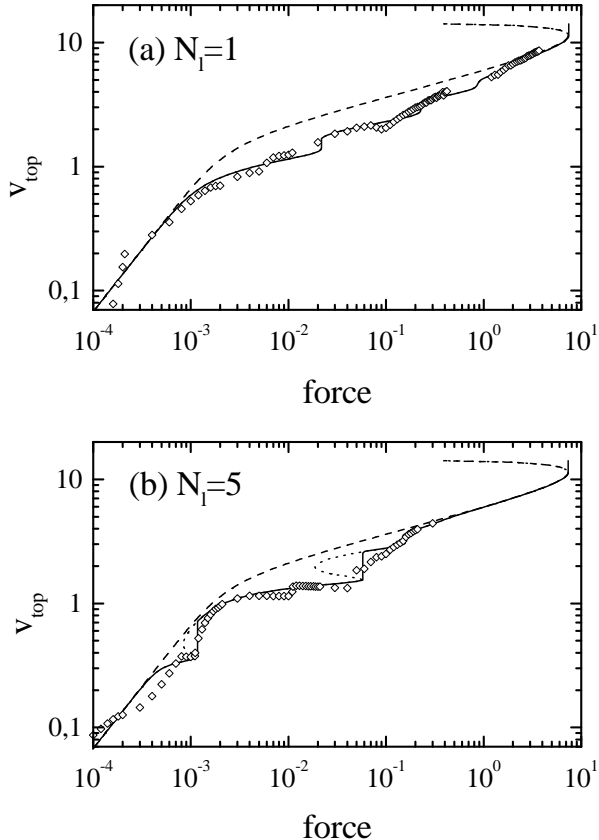}
\caption{
Fitting of the simulation results
for (a) $N_l=1$ and (b) $N_l=5$.
Open diamonds are for the simulation data,
the dashed curve shows the ``universal''
dependence (\ref{etaNum}),
the dotted curve is obtained with
Eqs.\ (\ref{res5}--\ref{res7}),
and the solid curve is obtained from the dotted one
by removing the unstable branches
where $dv_{\rm top}(f)/df<0$.
For the fitting we used the following frequencies,
$\omega_0^{(1)} = 7.3$,
$\omega_0^{(2)} = 5$,
$\omega_0^{(3)} = 3.3$,
$\omega_0^{(4)} = 1.65$,
$\omega_0^{(5)} = 0.4$,
and the following fitting parameters
for the widths of the resonance modes
and the oscillating forces.
(a) For $N_l=1$:
$\widetilde{\eta}_0^{(1)}=5$, $f_0^{(1)}=5$;
$\widetilde{\eta}_0^{(2)}=3$, $f_0^{(2)}=3$;
$\widetilde{\eta}_0^{(3)}=3$, $f_0^{(3)}=1.3$;
$\widetilde{\eta}_0^{(4)}=1.5$, $f_0^{(4)}=0.2$;
$f_0^{(5)}=0$.
(b) For $N_l=5$:
$f_0^{(1)}=0$;
$\widetilde{\eta}_0^{(2)}=3$, $f_0^{(2)}=0.9$;
$\widetilde{\eta}_0^{(3)}=1$, $f_0^{(3)}=0.7$;
$\widetilde{\eta}_0^{(4)}=0.3$, $f_0^{(4)}=0.2$;
$\widetilde{\eta}_0^{(5)}=0.1$, $f_0^{(5)}=0.006$.
}
\label{fig085}
\end{figure}

To fit the simulation data,
we also assumed that
the oscillator's damping coefficient
depends on the frequency according to the expression
$\eta_0^{(i)} (\omega) = \widetilde{\eta}_0^{(i)} 
\eta_{\rm ph}(\omega) / 
\eta_{\rm ph} \left( \omega_0^{(i)} \right)$,
where $\widetilde{\eta}_0^{(i)}$ is a constant.
Then, from the spectra of Fig.\ \ref{fig060}
we choose the following five frequencies:
$\omega_0^{(1)} = 7.3$
which can be associated with 
the $z$ vibrations of the substrates,
$\omega_0^{(2)} = 5$
which describes 
the lowest $x$ and $y$ modes of the substrates,
$\omega_0^{(3)} = 3.3$
which corresponds to the lubricant oscillations,
the frequency
$\omega_0^{(4)} = \frac{1}{2} \omega_0^{(3)}$
which describes the excitation of the lubricant
by the second harmonic of $\omega_{\rm wash}$,
and finally (for the $N_l=5$ case only)
$\omega_0^{(5)} = 0.4$
which may be associated with the vibration
of the top substrate as a whole.
The results of fitting
are presented in Fig.\ \ref{fig085},
where the values of the fitting parameters
$\widetilde{\eta}_0^{(i)}$
and
$f_0^{(i)}$
are given in the caption to the figure.
One can see that this simple approach gives
a reasonable agreement with the simulation data
and shows that the velocity of the top substrate
is determined by the ``perfect-sliding'' approximation
at forces $f<10^{-3}$,
by the intrinsic losses corresponded
to excitation of lubricant vibrations
at forces $10^{-3} <f< 0.1$,
and by excitation of the substrate vibrations
at higher forces.
However, we have to note that a general theory
of kinetic friction which could give, e.g.,
the values of the parameters
$\widetilde{\eta}_0^{(i)}$ and $f_0^{(i)}$
from first principles,
is still lacking.
%

%------------------------------------------------------------------
\subsection{Minimal sliding velocity}
\label{risk}

\begin{figure}
\epsfxsize=\hsize
\epsfbox{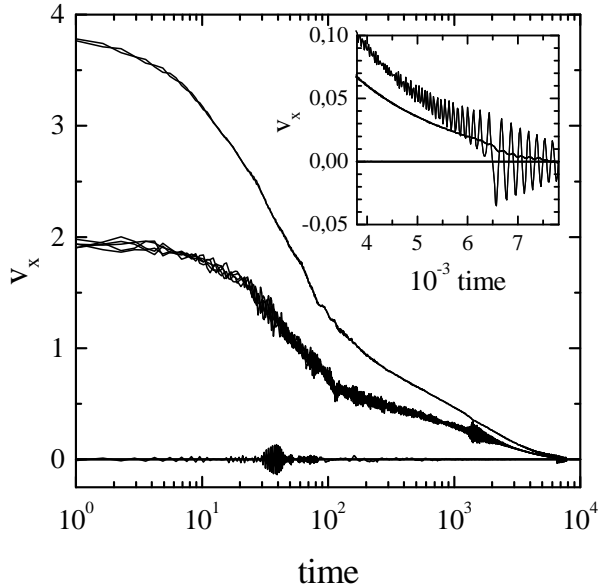}
\caption{
The $x$ velocities of the top substrate
and all layers (including the s-layers)
versus time for the ``free'' run,
when the five-layer system starts
from the steady state corresponded to
$f=0.2$ and then the driving force is removed.
Inset: enlarged part of the dependences
at late times.}
\label{fig090}
\end{figure}

As mentioned above, the system exhibits hysteresis:
when the force decreases starting from the sliding state,
a backward transition takes place at $f=f_b < f_s$,
and the velocity drops down from a finite value
$v_{\rm top}=v_b$ to zero.
The same drop of the velocity is observed in free runs,
when one starts with the solid-sliding steady state
and then remove the driving:
the velocity $v_{\rm top}$ slowly decreases till it
reaches the minimal value $v_b$ and then drops to zero
as shown in Fig.\ \ref{fig090}.
It is well known (e.g., see \cite{Risken}) that
a similar hysteresis is observed for a single driven particle at $T=0$.
If a particle of mass $M$ is placed into 
a sinusoidal external potential of (total) height
${\cal E}= \max V(x) - \min V(x)$
and period $a$,
and is driven by the dc force $F$,
the forward locked-to-running transition takes place at
$F=F_s=\pi {\cal E} / a$, and the backward transition, at
$F=F_b= \left( 2\sqrt{2}/ \pi \right) \eta \, \sqrt{M {\cal E}}$,
when the velocity is 
$\langle v_b \rangle \sim
\sqrt{a F_s /M}$.
In the underdamped case such that
$ \eta < \eta_c \equiv \left( \pi^2 / 2 \sqrt{2} \right) 
\sqrt{{\cal E}/{Ma^2}}$, we have $F_b<F_s$. Therefore
the system exhibits hysteresis due to the inertia of the particle.

If the top substrate with the attached s-layer
would be treated as a rigid object, 
taking $a=3$ and $m_S+m_s=2$, the critical damping would be
\begin{equation}
\eta_c =
\left[
\frac{\pi^3}{8} \,
\frac{f_s}{a (m_S+m_s)}
\right]^{1/2}
\approx 0.8 \, \sqrt{f_s}.
\end{equation}
The effective friction coefficient just above
the transition to sliding is
$\eta_s = f_s / (m_S+m_s) v_s$.
Hysteresis is expected if $\eta_s < \eta_c$, i.e.\ 
$v_s > v_{bR}$, where
\begin{equation}
v_{bR} = \left( \frac{8}{\pi^3} \, 
\frac{a f_s}{m_S+m_s} \right)^{1/2}
\approx 0.62 \, \sqrt{f_s}.
\label{vb}
\end{equation}
In the simulations the transition to sliding occurs for 
$f_s \sim 10^{-3} - 10^{-2}$
while $v_s \sim 1$,
so, for the chosen set of parameters,
the system under investigation is clearly in
the underdamped limit. We expect
the backward transition at
$f_b=\left( 2 \sqrt{2}/\pi \sqrt{\pi} \right)
\eta_{\rm eff} \; \sqrt{(m_S+m_s) \; a f_s}
\approx 1.24 \, \eta_{\rm eff} \sqrt{f_s}$.

Now we can check whether the condition
$v_{bR} \approx 0.62 \, \sqrt{f_s}$
agrees with the simulation data.
For the solid-sliding regime with $V_{ll}=1$ at $T=0$ 
we found from free runs that
for $N_l=1$,
$f_s \approx (1.9 - 2)\!\cdot\!10^{-2}$,
$v_b \alt 8.5\!\cdot\!10^{-2}$, while
$v_{bR} \approx 8.7\!\cdot\!10^{-2}$;
for $N_l=2$,
$f_s \approx (8.9 - 9.1)\!\cdot\!10^{-3}$,
$v_b \alt 6\!\cdot\!10^{-2}$, while
$v_{bR} \approx 5.9\!\cdot\!10^{-2}$;
for $N_l=3$,
$f_s \approx (4 - 4.1)\!\cdot\!10^{-3}$,
$v_b \alt 5\!\cdot\!10^{-2}$, while
$v_{bR} \approx 4\!\cdot\!10^{-2}$;
and for $N_l=5$,
$f_s \approx (1.7 - 1.8)\!\cdot\!10^{-3}$,
$v_b \alt 3\!\cdot\!10^{-2}$, while
$v_{bR} \approx 2.6\!\cdot\!10^{-2}$.
We also made runs for a smaller value of the interaction
parameter $V_{ll}=1/3$, when the lubricant is weaker,
and found that
for $N_l=1$,
$f_s \approx (1.3 - 1.4)\!\cdot\!10^{-2}$,
$v_b \alt 8\!\cdot\!10^{-2}$, while
$v_{bR} \approx 7.2\!\cdot\!10^{-2}$;
and for $N_l=5$,
$f_s \approx (1 - 1.1)\!\cdot\!10^{-3}$,
$v_b \sim 1.5\!\cdot\!10^{-2}$, while
$v_{bR} \approx 2\!\cdot\!10^{-2}$.

Thus, this simple inertia consideration quite well describes
the hysteresis of the solid lubricant film
and, at the same time, it shows that
the minimal velocity of the top substrate
in the solid-sliding regime cannot be smaller than
$v_b \sim 10^{-2} \; n.u. \sim 10 \; m/s$,
which is to be compared with the experimental value
of the sliding to stick-slip transition
$v_c \sim 10^{-6} \; m/s$. This very large discrepancy will be
discussed in section \ref{discussion}.
On the other hand, the friction $\mu=|f/f_{\rm load}|$
in the solid-sliding regime may be as small as
$\mu \sim 10^{-4} - 10^{-2}$.

Note that the value of $v_b$ in Eq.\ (\ref{vb})
depends on the substrate mass $m_S$ 
which, in principle, may be taken arbitrary large. 
However, in fact only one or few boundary layers
of the substrate play a role,
and the results presented above remains valid,
at least qualitatively,
for the infinite substrates as well
\cite{P1994,Braun-new}.

%------------------------------------------------------------------
\subsection{Friction-induced melting}
\label{induced}

\begin{figure}
\epsfxsize=\hsize
\epsfbox{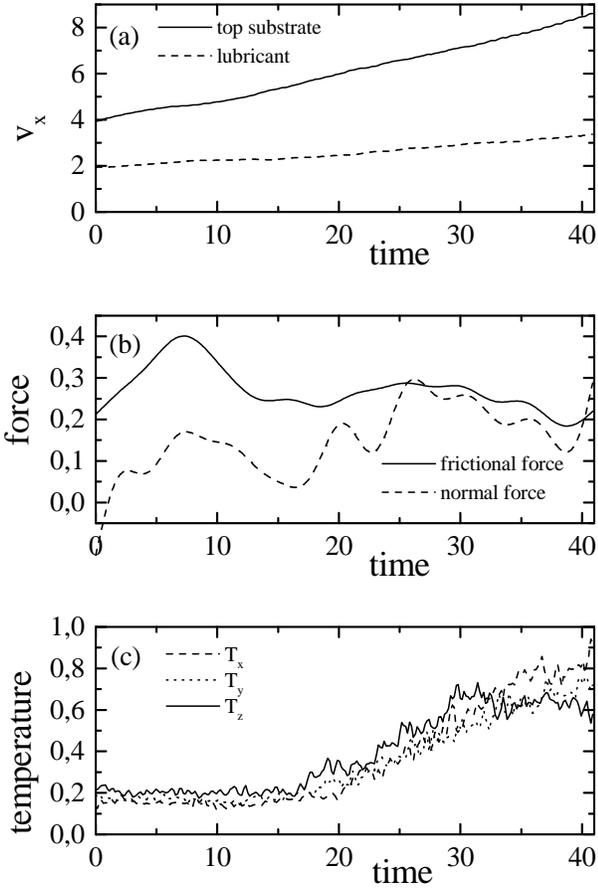}
\caption{
``Forced'' transition:
the dependences of 
(a) the $x$ velocities of the top substrate and the lubricant,
(b) the normal and frictional forces, and
(c) the lubricant temperatures
versus time, when the force $f=0.5$ is applied to an
initial configuration
corresponded to the $f=0.2$ steady state
of the five-layer film at $T=0$.}
\label{fig100}
\end{figure}

\begin{figure}
\epsfxsize=\hsize
\epsfbox{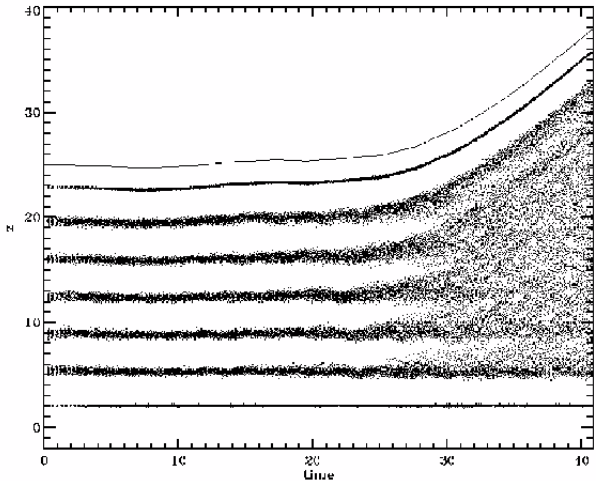}
\caption{
The $z$ coordinates of all atoms
and the top rigid substrate during the ``forced''
transition shown in Fig.\ \ref{fig100}.}
\label{fig102}
\end{figure}

When the driving force approaches the threshold value $f=f_f$, the
energy pumped into the lubricant due to sliding
can no longer be removed from the contact region
into the substrates, and the lubricant temperature strongly rises
up to its melting temperature. The solid-sliding regime is destroyed.
To study this process,
we started from the $f=0.2$ steady-state configuration
of the five-layer film,
which is close to the threshold value $f_f \approx 0.31$,
and applied the dc force $f=0.5$ to speed up the transition.
The results are shown in Fig.\ \ref{fig100}.
The velocity $v_x$ of the top substrate
grows approximately linearly with time.
Because the top rigid substrate with one substrate layer 
attached to it moves as a whole during this process,
we can use Newton's equation
\begin{equation}
(m_s+m_S) \dot{v}_{\alpha}
=f_{\alpha}^{(\Sigma)}
=f_{\alpha}^{\rm (intrinsic)} +
 f_{\alpha}^{\rm (external)},
\end{equation}
which for the $x$ coordinate takes the form
$2 \dot{v}_x =0.5 -f_{\rm fric}$,
where $f_{\rm fric}$ is the frictional force
acting on the top substrate from the lubricant.
One can see from Fig.\ \ref{fig100}b that $f_{\rm fric} \approx 0.3$
is the maximum force that lubricant losses can accommodate.
Similarly one may find the normal force
$f_P$ acting on the top substrate from the lubricant,
$2 \dot{v}_z =f_{\rm load} +f_P$,
which yields $f_P\approx 0.1$ 
(so that $f_{\rm load} + f_P \approx 0$)
at the beginning,
but at $t>20$ the force $f_P$ begins to grow
and the width of the lubricant begins to increase.
The lubricant temperature
(see Fig.\ \ref{fig100}c) which was equal to
$T_{\rm lub} \approx 0.2$ for the $f=0.2$ steady state,
at $t>20$ begins to grow up to
$T_{\rm lub} \sim 0.6$ -- $0.8$
which is much higher than the melting temperature
$T_{\rm melt} \approx 0.4$,
so the lubricant melts.
Details of the forced transition are shown in 
Fig.\ \ref{fig102},
where we plot the $z$ coordinates of all atoms in the system.
For $t<20$ the lubricant slides as an ideal solid
as it was for the $f=0.2$ steady state.
At $t\sim 20$ the upper lubricant layers begin
to slide over the lower layers, although
the layers still keep the ideal triangular structure.
At $t\sim 30$ the layers start to mix.
At $t\sim 40$ the lubricant is already 
in the melted state, and
the distribution of velocities across the lubricant
$v(z)$ exhibits a gradient.
After the melting, the lubricant cannot anymore
support the applied dc force $f=0.5$, and 
$v_{\rm top} \rightarrow \infty$.


\begin{figure}
\epsfxsize=\hsize
\epsfbox{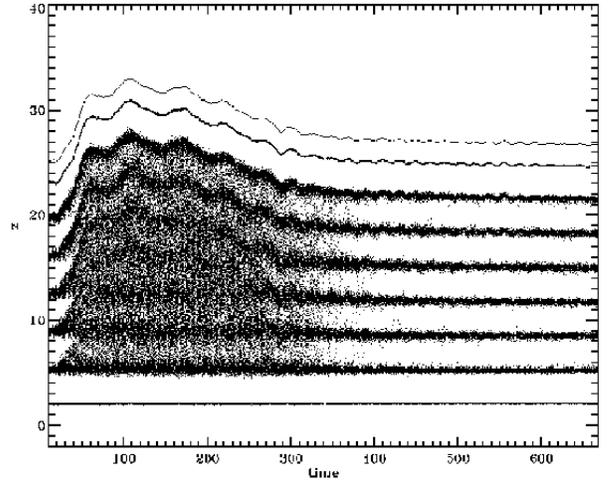}
\caption{
The $z$ coordinates of all atoms, when
the driving force is removed at the time moment
$t=10.25$
of the ``forced'' transition 
of Fig.\ \ref{fig102}.}
\label{fig104}
\end{figure}

In another simulation we turned off the force
at time $t=10.25$,
when the lubricant was still in the solid-sliding state,
and allowed the system to relax.
This situation mimics that of real systems, where
the driving force exists due to elastic stress
within the substrates.
Thus, when the relative velocities of the substrate
and the lubricant sharply increase,
the stress is relaxed and the driving force decreases
(in principle this situation can easily be modeled
with a spring attached to the top substrate
\cite{Robbins2}; 
unfortunately, one has to introduce simultaneously
few more model parameters).
In simulation we observed that although $f=0$ at $t>10.25$,
the top substrate continues to slide owing to its inertia.
During this sliding the lubricant again melts
as one can see from Fig.\ \ref{fig104}.
Then the lubricant temperature decreases, and
the lubricant solidifies into a six-layer
(amorphous) film at $t \agt 300$
when $T_{\rm lub} \sim 0.3$, and finally stops
(note that the final configuration obtained with this
force-induced melting and then freezing, is similar to
the configuration obtained by increasing temperature
till melting and then sharp decreasing it back to zero
without the driving).
The frictional force is large at the beginning
when the lubricant melts ($f_{\rm fric} \sim 0.15$),
but soon it quickly decreases to a value
$f_{\rm fric} \sim 0.01$.

We also made similar experiments but with an applied
force $f\neq 0$ ($f<0.5$) at  time  $t=10.25$.
We observed that, if
$f<0.005$, the system stops in a six-layer state.
For $0.005 <f< 0.03$ we observed a sliding
of the ``frozen'' six-layer lubricant;
the lubricant temperature during the sliding is
$T_{\rm lub} \approx 0.06$ at $f=0.005$,
$T_{\rm lub} \approx 0.14$ at $f=0.01$, and
$T_{\rm lub} \approx 0.19$ at $f=0.02$.
For $f=0.03$ we observed that $v\rightarrow \infty$,
and the lubricant severs in the middle. From the slope 
of the $v_{\rm top} (t)$ dependences
we found that $f_{\rm fric} \approx 0.028$,
thus the frozen lubricant cannot support
a driving force larger than $f \approx 0.03$.

Comparing the results obtained for a realistic external
damping with the ones described above in the introduction,
where the model with a constant damping coefficient was used, 
we see the essential difference.
First, in the latter case the lubricant melts
immediately when it begins to slide, i.e.\ $f_f < f_s$.
This difference is due to the parameters chosen
for the interatomic interaction.
When the interaction of the lubricant with
the substrates is much larger than the interaction within
the lubricant, the scenario reviewed in the introduction is observed
(the same also occurs if the lubricant is commensurate
and perfectly aligned with the substrates).
Second, after the shear-induced melting,
the layer-over-layer sliding was observed,
while now the transition to a ``gas'' phase takes place.
Third, the back transition with force decrease now proceeds into
an amorphous (frozen) state and not into the ideal solid one.
These last two differences are due to damping mechanism
used in the simulation. 
As will be shown below in Sec.\ \ref{discussion},
when the model with a constant damping is used,
the scenarios of the previous works are observed, which points out the
importance of the damping mechanism on some results.

%------------------------------------------------------------------
\subsection{Sliding of curved substrates}

The results presented above, were obtained
for ideally flat substrates. Due to periodic
boundary conditions we have in fact infinite
substrates and the lubricant atoms are strictly confined
between them.
In a real system, however, the lubricant may
leave the contact region through open boundaries.
Besides, the substrates may be not
ideally flat.
To study this situation, we made a few simulations
for curved substrates for the $N_l=5$ system.
The results are the following.

\begin{figure}
\epsfxsize=\hsize
\epsfbox{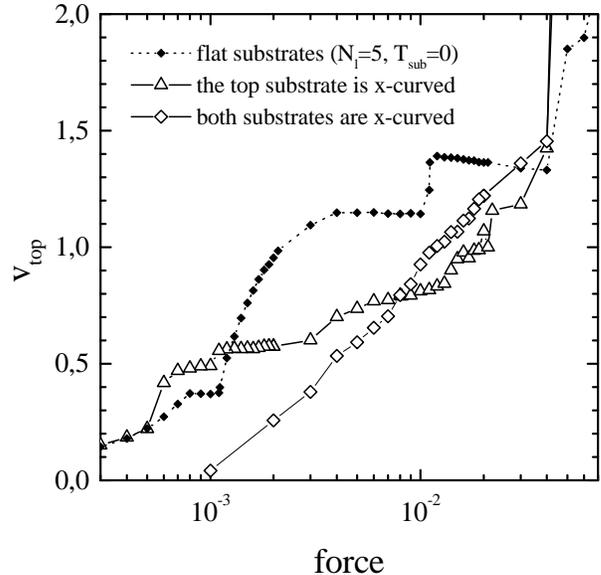}
\caption{
The $x$ velocity of the top substrate versus force
for the five-layer lubricant film
when the top rigid substrate is curved
($h_x^{\rm (up)}=1$,
$h_y^{\rm (up)}=
h_x^{\rm (dn)}=
h_y^{\rm (dn)}=0$,
the up triangles)
and when both top and bottom rigid substrates are curved
($h_x^{\rm (up)}=
h_x^{\rm (dn)}=0.5$,
$h_y^{\rm (up)}=
h_y^{\rm (dn)}=0$,
the down triangles).}
\label{fig110}
\end{figure}

The $v_{\rm top}(f)$ dependence 
for the case when only one of the substrates 
(e.g., the top one) is curved in the $x$ direction
(along the driving direction) is presented in
Fig.\ \ref{fig110} by up triangles.
In this case the lubricant film is approximately flat
and follows the bottom (flat) substrate.
At very low forces, $f \leq 5\!\cdot\! 10^{-4}$,
the lubricant moves together with the top substrate.
For forces within the interval
$5\!\cdot\! 10^{-4} <f< 0.01$
the ideal solid-sliding regime exists.
The lubricant film is approximately flat and
slides over the flat bottom substrate,
and the top substrate slides over the lubricant;
in the contact region the lubricant is slightly compressed.
For forces within the interval
$10^{-3} <f< 10^{-2}$
the velocity is approximately two times
smaller than it was for the flat substrates.
Note that because the real contact area takes now only
half of the total surface, the real load pressure
in the contact is two times larger.
At higher forces, $f>10^{-2}$, the perturbation
of the lubricant becomes significant.
At $f \agt 0.011$ some atoms from the topmost
lubricant layer begin to escape into the empty space
between the lubricant film and the curved top
substrate, although they are pushed back
when the bottom of the curved top substrate
reaches them during the sliding. This mixing increases with 
increasing of the dc force, and
at $f \agt 0.02$
the lubricant fills the empty space.
In the thick part there are six layers.
Finally at $f\approx 0.05$ 
(i.e.\ much earlier than for the flat substrates)
the lubricant melts,
and the solid-sliding regime is destroyed.

When both substrates are curved
in the $x$ direction with the same curvature parameters,
the ground state of the system corresponds
to the five-layer lubricant film with an ideal
structure of the layers. The curvatures
of the substrates coincide,
and the lubricant film just follows the substrates.
But when the top substrate moves,
one  have alternatively a configuration
where the curvatures coincide as in the ground state,
and a saddle configuration 
where the curvatures anti-coincide
so that in one half of the contact area
the film is compressed, while in the other
half the distance between the substrates
is much larger than the width of the five-layer lubricant film.
Because a thin lubricant film cannot 
stay free in space but is attracted to one of the substrates,
it has to follow either the bottom or top substrate.
In simulation we observed that
the lubricant indeed takes a ground state-like configuration
when the substrate curvatures coincide, while
in the saddle points the lubricant follows
the top substrate at the left-hand side,
where the $z$-coordinate of the top substrate
goes up (so it ``drags'' the lubricant),
and at the right-hand side,
where the $z$-coordinate of the bottom substrate
goes up,
the lubricant follows the bottom substrate.
Thus, in the middle point the lubricant has to switch
between these two configurations,
and this  decreases the system mobility.
The $v_{\rm top}(f)$ dependence for this case
is shown in Fig.\ \ref{fig110} by down triangles.
The backward transition now takes place at
$f_b \approx (2-3)\!\cdot\! 10^{-3}$.
When the force increases up to $f \sim 0.04$,
the film does not have enough time 
to reach the five-layer ground state-like configuration
between the saddle points.
It first takes a six-layer structure
but soon (at $f_f \agt 0.04$)
the lubricant melts and the sliding is destroyed.
Note, however, that the situation described above corresponds
to a rare case when the curvature parameters of both substrates
are exactly equal one another. In all other cases,
in a given region only one of the substrates
is usually strongly curved.

We also checked the system mobility
when the substrates are curved in the $y$ direction
(perpendicular to the sliding direction).
When only the top substrate is curved,
the velocity remains approximately the same
as that for the flat substrates
(for example, for $f=2\!\cdot\! 10^{-3}$
the velocity decreases from $v\approx 0.98$
for the flat substrates
to $v_{\rm top} \approx 0.96$ for the curved top substrate).
When both substrates are curved,
the velocity remains unchanged
(e.g., for $f=2\!\cdot\! 10^{-3}$
the velocity increases to $v_{\rm top} \approx 0.985$).
Finally, when the top substrate is curved
in both $x$ and $y$ directions simultaneously,
the velocity decreases for the 
$f=2\!\cdot\! 10^{-3}$ case
to the value $v_{\rm top} \approx 0.52$
(compare with $v_{\rm top} \approx 0.57$ for the case
when the top substrate is curved
in $x$ direction only).

Thus, we may conclude that the simulation results
obtained for the system with flat substrates stay valid, at least
qualitatively, for curved ones.

%==================================================================
\section{Discussion and conclusions}
\label{discussion}

To summarize, we have studied the friction for a thin solid lubricant
film between two flat solid substrates and found that
the frictional losses are mainly due to the excitation of $z$ vibrations
in the lubricant with the washboard frequency.
The anharmonicity of lubricant vibrations is found to be
important and leads to heating of the lubricant.
Thus, a first lesson learned from the simulation of a complex
model, is that simple FK-type models with harmonic springs between
the atoms cannot describe the kinetic friction properly.

\begin{figure}
\epsfxsize=\hsize
\epsfbox{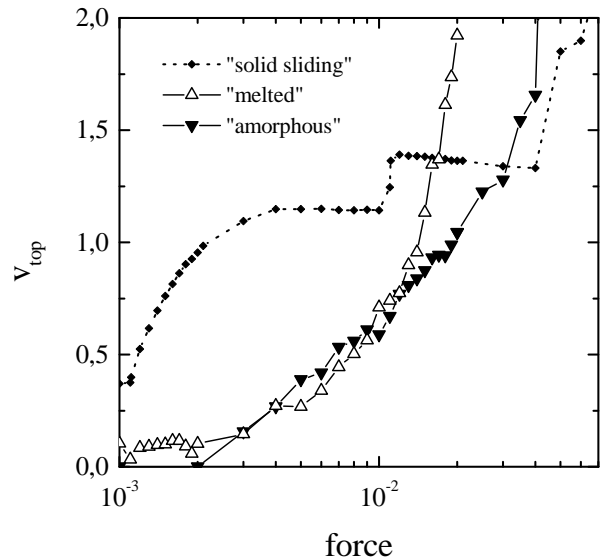}
\caption{
The $v_{\rm top}(f)$ dependences for the
melted lubricant 
($T_{\rm lub} \approx 0.5$, the up triangles)
and the frozen (amorphous) lubricant
(the down triangles).}
\label{fig120}
\end{figure}

The solid lubricant between two flat 
(or at least not too rough) substrates
provides %, probably 
the smallest possible frictional force
for high-speed systems. 
The energy losses are very small both in the low-velocity
(but $v_{\rm top} > v_b \sim 10 \; m/s$)
case, as well as in the high-velocity
($v_{\rm top} \agt v_{\rm sound}$) regime,
the later regime being stable if
there is a gap between the lubricant 
and substrate phonon spectra.
If the lubricant film is in a liquid state
(melted either due to high substrate temperature
or because of very high speed),
the frictional force is much larger as seen from Fig.\ \ref{fig120}.
When the lubricant is frozen back from the melted state,
it takes an amorphous structure because, due to the contact
with the substrates, the energy is removed from the lubricant
very fast, and the confined lubricant film 
has no time to order.
The frictional force in the case of an amorphous lubricant
may be even larger than for the melted one (see Fig.\ \ref{fig120}).

The solid lubricant between flat substrates
leads also to quite small values of the static frictional force $f_s$.
This agrees with the results of previous simulations
(see \cite{Robbins2} and references therein).
The new result of the present work is the dependence
of $f_s$ on the number of layers in the solid lubricant film.
The exponential decrease of $f_s$ with $N_l$
may be used for explanation of experimental values
of the velocity of the transition from smooth sliding
to stick-slip motion as will be discussed below.

We also checked how the results are sensitive to the
parameters of the model.
First, changing the mass $m_S$ of the top rigid substrate
practically does not change the adiabatic $v_{\rm top}(f)$ dependence,
thus the velocity in the steady solid-sliding regime
does not depend on $m_S$.
Another important parameter is the minimal external damping
$\eta_{\rm min}$ which models the multi-phonon and/or ``electron-hole''
contributions to the energy exchange of lubricant with substrates.
We found that the results are approximately unchanged
for $v_{\rm top} \agt 1$, when 
the energy losses inside the lubricant play the main role.
At small forces/velocities,
when $v_{\rm top} \alt 1$, 
the values of $v_{\rm top}$ and $z_{\rm top}$
are larger for smaller values of $\eta_{\rm min}$
and agree with the ``perfect-sliding'' dependence (\ref{etaNum}).

\medskip
Finally, let us discuss the relation of the microscopic simulations
of the present work to macroscopic friction.

\begin{figure}
\epsfxsize=\hsize
\epsfbox{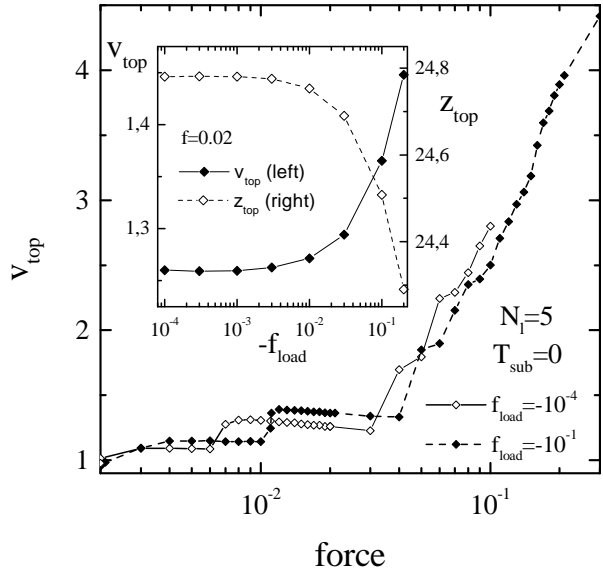}
\caption{
The dependences $v_{\rm top}(f)$
for two values of the load
$f_{\rm load}=-0.1$ (dashed curve and solid diamonds)
and
$f_{\rm load}=-10^{-4}$ (solid curve and open diamonds)
for the five-layer film at $T=0$.
Inset:
$v_{\rm top}(f_{\rm load})$ (left axes)
and
$z_{\rm top}(f_{\rm load})$ (right axes)
for the fixed value of the driving force
$f=0.02$.}
\label{fig130}
\end{figure}

\medskip
{\it Dependence on load.\/}
We varied $f_{\rm load}$ to test the Amontons first 
law $F_{\rm fric} =\mu F_{\rm load}$,
where $\mu$ (called the friction coefficient by tribologists)
should be approximately constant.
For example, the simulation results for a much lower load
$f_{\rm load}=-10^{-4}$
than that used in all simulation presented above,
are shown in Fig.\ \ref{fig130}.

The trivial result is that the width of lubricant
increases with decreasing of $f_{\rm load}$.
Therefore, the external damping coefficient,
which exponentially depends on $z$, decreases with $f_{\rm load}$.
As a result the backward transition now takes place at
a lower force $f=f_b \approx (3-4)\!\cdot\!10^{-5}$ when
$v_b \alt 3\!\cdot\!10^{-2}$.
The forward transition also occurs earlier, at
$f_f \approx 0.1 - 0.11$ when $v_b \alt 3$.
For $v_{\rm top} \agt 1$ the dependence $v_{\rm top}(f)$
is approximately the same as that for the high load,
while for lower forces when $v_{\rm top}<1$,
the values of $v_{\rm top}$ for the same values of $f$
are larger than those for the $f_{\rm load}=-0.1$ case,
because the external friction is smaller 
due to larger values of $z$.
The dependences of $v_{\rm top}$ and $z_{\rm top}$ on $f_{\rm load}$
for a fixed value of the dc force $f=0.02$
are shown in inset of Fig.\ \ref{fig130}.
Surprisingly, $v_{\rm top}$ is not decreasing but increases
when $f_{\rm load}$ increases, because for the chosen value
of the dc force, a resonance effect plays a role
(the washboard frequency penetrates into the lubricant phonon spectrum).
In general, however, the variation of $f_{\rm load}$
results in a little change only. This confirms once more that
the empirical Amontons first law works due to change 
of the contact area with load \cite{BT1986}.
On the contrary, He and Robbins \cite{HR2000}
have found that the frictional force is directly proportional to the load.
However, their simulation has been done with
the const-$v$ algorithm at low velocities
($v_{\rm top} < 0.1 \; n.u.$)
and much higher temperatures than studied in our simulations.
Besides, the authors of \cite{HR2000} used the Langevin equations
with a constant external damping.

\medskip
{\it Dependence of results on the interaction parameters.\/}
Above we have studied the system behavior
for the fixed set of parameters,
when the interaction within the lubricant,
$V_{ll}=1$,
is much larger than the interaction
of the lubricant with substrates,
$V_{sl}=1/3$.
For these parameters the solid-sliding regime
had indeed to be expected.
Below we present a few simulation results 
for other sets of parameters.

\begin{figure}
\epsfxsize=\hsize
\epsfbox{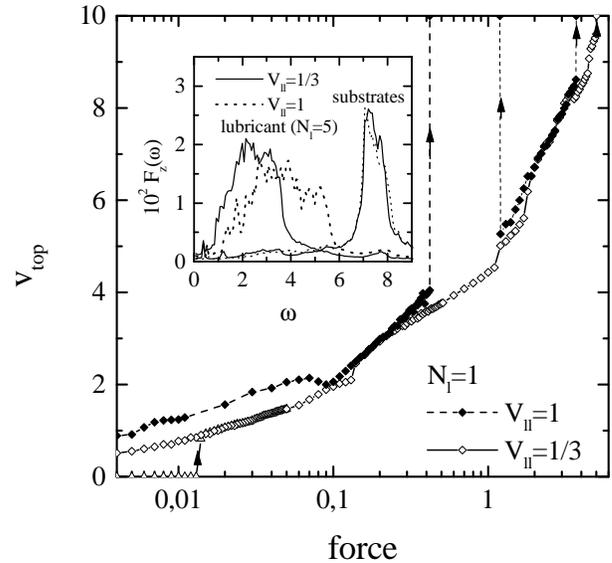}
\caption{
The $v_{\rm top}(f)$ dependence
for the one-layer lubricant with
the interatomic interactions
$V_{ll}=1/3$ (solid curve and open diamonds)
comparing with that
for the $V_{ll}=1$ case (dashed curve and solid diamonds).
Inset: the $V_{ll}=1/3$ (solid curves)
and $V_{ll}=1$ (dotted curves) spectra
for the five-layer lubricant.}
\label{fig140}
\end{figure}

{\it First\/}, let us consider the case with
$V_{ll}=V_{sl}=1/3$,
when the interaction within the lubricant has the same magnitude
as the lubricant--substrate interaction.
In fact, this case is similar to that studied above,
because an effective lubricant--substrate interaction
is again smaller than the interaction within the lubricant
due to mismatch of the substrate/lubricant lattice constants.
However, an essential new feature of the present case
is that now the phonon frequencies of the lubricant are smaller.
Because the interaction within the substrates
and within the lubricant differs by nine times,
the substrate and lubricant phonon spectra are
well separated one from another
even for the five-layer lubricant film
as shown, for example,
in inset of Fig.\ \ref{fig140} for the $z$ component.
There is a wide gap for
the frequencies $4 < \omega < 6.5$
where the density of phonon states is very small.
Thus, one could expect that if the washboard
frequency comes into this frequency interval,
the effective friction should be small,
and the solid sliding regime will persist up to very high velocities.
Indeed, the simulation results for the one-layer lubricant
film presented in Fig.\ \ref{fig140}, show that
the solid-sliding regime indeed survives now
till the very high force
($f_f \agt 5$) and velocity ($v_f\approx 10$).
This effect is similar to the ``reentrant'' transition
observed by Stevens and Robbins
\cite{SR1993},
where with the increase of the driving velocity
the lubricant melts but then, at a higher velocity,
it solidifies again.
In our simulation with a small external damping
we did not obtained melting but instead observed
a large increase of the lubricant temperature.
For velocities $v_{\rm top} > 3$
(recall that in the previous case the solid-sliding regime
was already destroyed at so high velocities)
the washboard frequency is higher than
the lubricant phonon frequencies,
thus the excitation of oscillations in the lubricant as well as
the effective lubricant temperature decrease.
This regime corresponds to the ``flying'' one, 
i.e.\ the decoupling of phonon spectra of the lubricant
and the substrates stabilizes the ``flying'' state.
With increasing dc force
the lubricant remains sliding in a solid state,
while the amplitude of the $z$ oscillations
of the substrate atoms strongly increases at $v_{\rm top} > 6$, 
and this finally destroys the solid-sliding regime.
Note that for the forces $0.1<f<0.4$, 
when $2 < v_{\rm top} < 3$,
the system behavior is similar to that described above, 
because the densities of lubricant
vibrations are similar at these washboard frequencies.
At smaller forces, $f<0.1$, the effective friction
as well as the lubricant temperature are larger
for the weak-lubricant case than those for the previous case.
The backward transition to the locked state
takes place now at $f_b \alt 0.003$, when $v_b \approx 0.15$.
The static frictional force,
$f_s \approx 0.013 - 0.014$, 
is also smaller than that for the previous case
(recall $f_s \approx 0.02$ for the $V_{ll}=1$ case).
Note, however, that we observed the ``flying'' regime only for
the one-layer film. For the five-layer lubricant
the forward transition takes place at
$f_f \approx 0.025 - 0.026$, i.e.\ much earlier than
for the $V_{ll}=1$ case,
and the attempts to get a high-velocity regime failed.
Note that although it seems quite unusual to have a velocity
close to the sound speed in real machines,
the ``flying'' regime may be reached in, e.g.,
next generation of hard-disk drives. In a disk with a radius of
4~cm and a rotating speed of 10000~rpm, the speed of the head with
respect to the plateau already exceeds 40~m/s, i.e. the crash of the
head on the plateau in a failure is an example of high speed friction,
although, unfortunately, the ``flying'' regime would not be reached
the current generation of hard disks.

\begin{figure}
\epsfxsize=\hsize
\epsfbox{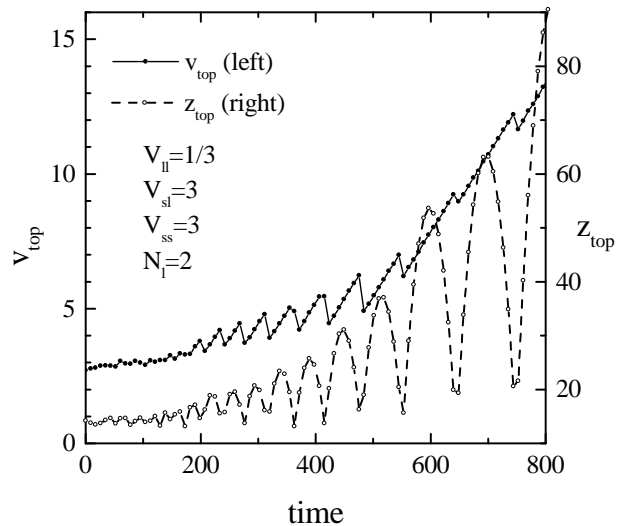}
\caption{
Beginning of motion for the ``glued'' two-layer lubricant film
($V_{ss}=V_{ls}=3$, $V_{ll}=1/3$).}
\label{fig150}
\end{figure}

{\it Second\/},
we studied the opposite case,
when the lubricant--substrate interaction is much larger
than the interaction within the lubricant, namely
$V_{sl}=3$ and $V_{ll}=1/3$.
In this case the two utmost lubricant layers
tend to be attached to the corresponding substrates,
so a sliding can only emerge inside the lubricant.
In fact, this situation is close to real oil lubricant,
where the interaction with substrates prevents
the lubricant from being squeezed out from the contact areas.
In the case of $N_l=2$ all lubricant atoms are attached
by the substrates, so there are no ``lubricant''
between the two sliding ``bodies'', each consisting of
the rigid substrate with one attached s-layer
and one ``glued'' layer of l-atoms.
Thus, the situation is close to the dry friction
for two commensurate perfectly aligned substrates.
However, the interaction between the bodies, $V_{ll}=1/3$,
is much smaller than the interaction inside them,
where $V_{ss}=V_{sl}=3$.
In this case at some critical force,
$f_s \sim 0.08$,
the bodies decouple one from another,
the distance between them increases,
and they begin to ``fly'' almost without friction.
But the loading force pushes the bodies into contact again,
so they collide and almost elastically repulse,
then again collide, {\it etc\/}.
The motion of the top substrate resembles a series of jumps
as shown in Fig.\ \ref{fig150}.
If the driving force remains constant,
the amplitude of jumps increases with time,
and finally at one collision the surface lubricant
layers are destroyed.

\begin{figure}
\epsfxsize=\hsize
\epsfbox{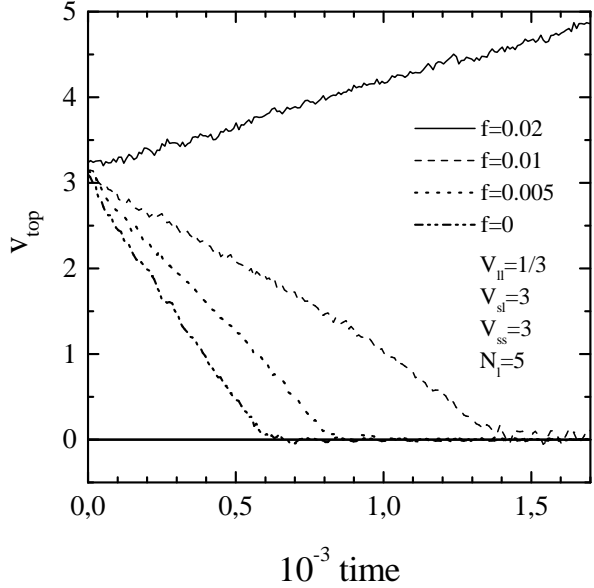}
\caption{
Motion of the ``glued'' five-layer lubricant film
($V_{ss}=V_{ls}=3$, $V_{ll}=1/3$)
for different driving forces.}
\label{fig155}
\end{figure}

For a thicker lubricant film, $N_l>2$,
there are weakly interacting lubricant atoms between the two bodies.
Note that for these parameters in the ground state configuration
the utmost lubricant layers which are attached to the substrates,
have slightly more atoms than the middle layer(s).
When the motion begins at $f=f_s$ in this case
(e.g., for the $N_l=3$ or $N_l=5$ system $f_s \sim 0.03 - 0.07$),
the middle lubricant layers melt according to a 3D melting mechanism
(one layer in the $N_l=3$ system or three layers for the $N_l=5$ case),
and the width of the lubricant film increases.
Starting with the configuration just after the beginning of the motion
(e.g., that for $f=0.08$ for the $N_l=5$ system),
we abruptly decreased the dc force to a smaller value.
The results are presented in Fig.\ \ref{fig155}.
If $f \leq 0.01$, the motion locks in the five-layer solid state,
while for $f \geq 0.02$ the top body is ``flying'' and
its velocity increases linearly with time.
Thus, the frictional force for a melted lubricant
in this system is $f_{\rm fric} \sim 0.015$.

{\it Third\/},
all the scenarios described above were obtained for the model where
the external damping depends on the distance
from the substrates and, thus, it is very small
for the middle lubricant layer(s).
However, if we change the model in the way that
the minimal external damping would not depend
neither on distance nor on velocity,
the scenario changes to that already observed
in the previous works \cite{TR1990a,BBR99}
and described in the Introduction.
Now when the sliding begins the width does not
increase significantly and the sliding steady state survives
with increasing the dc force.
% as shown in Fig.\ 356.opj/Graph1 for the $N_l=3$ system.
With  increasing force, the middle layers melt due to sliding
according to a 2D melting mechanism,
i.e.\ the lubricant keeps its layered structure
and different lubricant layers slide one over another.
This model may be applied to lubricants with complex
molecules like organic ones, where an internal viscosity
of the lubricant plays an important role.

\bigskip
{\it The transition from smooth sliding to stick-slip motion.\/}
If, instead of a constant dc force we would drive 
the system through a spring attached to a stage 
moving with the velocity $v_{\rm top}$,
the system would exhibit a
transition from smooth sliding to stick-slip motion
at the velocity $v_{\rm top} \approx v_b$
\cite{Braun-new}.
A similar result would be
expected with a large enough substrate block so that its elasticity
would start playing a role. Experimentally
the transition from smooth sliding to stick-slip motion
is observed in almost all systems
where the static friction force is nonzero
(the zero $f_s$ corresponds to liquid lubricant).
There are at least three mechanisms that can explain
the transition from smooth to stick-slip motion:
(a) ``inertia'' mechanism,
when the backward sliding-to-locked transition
occurs at the force $f_b < f_s$ due to inertia
of the moving lubricant in the underdamped system
(this mechanism was discussed above in Sec.\ \ref{risk});
(b) ``melting--freezing'' mechanism \cite{TR1990a,P0,P00,Persson},
when the lubricant undergo dynamical phase transitions between
a fluidized state during slip (sliding)
and a solid state during stick; and
(c) ``memory'' mechanism (e.g., see \cite{P0,P00,Gen}),
when, after the sliding-to-locked transition,
the static friction force increases with the time of stationary contact.
The simulation results of the present work show that 
the ``inertia'' mechanism
leads to the critical velocity 
$v_c \sim 10^{-2} \; n.u. \approx 10 \; m/s$,
and we do not see any reason to get a much lower value for $v_c$. 
This value is orders of magnitude larger than the experimental value,
which rules out the inertia mechanism to explain the experimentally
observed stick-slip to smooth sliding transition.
For the model parameters used in the present work,
we also did not observed the melting of thin lubricant films
except for extremely large forces and velocities.
The lubricant may melt at $f>f_s$ if
it is strongly coupled with the substrates
or, in a more specific case, when the lubricant is perfectly 
commensurate and aligned with the substrates
\cite{TR1990a,Braun-new}.
But, in such situations, when the driving force is reduced,
the lubricant quickly solidifies again
due to a fast energy exchange with the substrates
(as discussed above in Sec.\ \ref{induced}), and
the backward transition to the locked state takes place
at the velocity $v_c \sim 1 \; n.u.$
\cite{TR1990a,P1994,Braun-new}.
Thus, in this case too the top substrate cannot move smoothly 
with a low velocity at the microscopic scale.
In experiments the transition 
from stick-slip to smooth motion
is observed usually at a velocity
$v_c \sim 1 \; \mu m/s \approx 10^{-9} \; n.u.$ 
\cite{Mate95}.
Even if we suppose that there are only few contacts
(asperities) between the solids
and only one of the contacts moves at a time, we do not see a way
by which the macroscopic stick-slip to smooth transition
could be explained by the microscopic (atomic-scale) mechanisms.
Thus, we conclude that the mechanism
responsible for the experimentally observed
stick-slip to smooth transition,
at least for ``simple'' lubricant molecules,
should be the ``memory'' effects.
After the sliding-to-locked transition,
the static friction force increases with time
due to plastic deformation of the lubricant,
for example,
because of an increase of the area of real contact,
or due to the decrease of the lubricant width
(according to the simulation results of the present work,
the squeezing of the lubricant leads to the increase of $f_s$),
or due to interdiffusion of lubricant molecules
between different layers of the lubricant
in the case of lubricant consisting of long-chain polymers.
All these processes involve plastic deformations
and thus are characterised by macroscopic-scale
characteristic times.
Then, the experimentally observed velocities of 
the transition from smooth sliding to stick-stick motion
can be explained with the help of earthquakes-like
models
\cite{BK1967,R1983,CLS1994}
as was demonstrated in
\cite{P00,P1995,P1997,Braun-new}.
Consequently the analysis of the order of magnitude of the stick-slip to
smooth-sliding transition velocity suggests that the explanation of
this penomenon should not be seeked at the microscopic scale and is
probably out of reach of current molecular dynamics simulations.
Besides, this also explains why the experimentally measured
kinetic friction in the smooth sliding regime
is often almost independent of the velocity,
while the microscopic one shows a very strong dependence.
Indeed, if the macroscopic smooth sliding
corresponds to the microscopic stick-slip regime,
the macroscopic kinetic friction would correspond
to microscopic static friction force which is constant. Therefore the
evaluation of the orders of magnitude that we made in this work lead
us to conclude that there is a fundamental difference between
microscopic friction and the macroscopic experimental properties. This
should perhaps be considered for the current projects to build
micro-machines. 

\bigskip
Finally, in real experimental systems the rate of change
of the driving force corresponds to adiabatically slow
variation of $f$ in simulations.
Therefore, to obtain reliable results, which do not 
depend on an initial configuration,
it is important to reach the steady state for a given $f$.
As observed in the present work,
lubricant systems are often characterized by quite long
relaxation times
$\tau_{\rm relax} > 10^3 \; n.u. \approx 10^{-10} \; s$
(or even $\tau_{\rm relax} \gg 10^{-10} \; s$
in the solid-sliding regime at velocities
$v_{\rm top} < 10 \; m/s$).
In simulations for large systems with realistic interactions
one could expect to reach the stationary state for one given
set of parameters (including the value of the driving force)
only, and not for adiabatically changing force.
Thus, although being very attractive, the ``millions of atoms'' simulations
hardly could give reliable values for the kinetic friction in their
present stage.
This is why we believe ``simple'' models are still very important.
But they must also be taken with great caution because,
as we have shown in the present work,
the results obtained with simple models
may also strongly depend on the model chosen, in particular of the
damping introduced in the simulations. We tried here to introduce a
damping mechanism motivated by microscopic considerations and we
showed that the results are sometimes very different from those given
by a uniform damping coefficient as often considered.

%==================================================================
\acknowledgments
We wish to express our gratitude to
A.\ R.\ Bishop,
B.\ N.\ J.\ Persson,
J.\ R\"{o}der, and
M.\ Urbakh
for helpful discussions.
This research is supported in part by
NATO grant No.\ HTECH.LG.971372.
O.B.\ was also partially supported by
INTAS Grant No.\ 97-31061.

%==================================================================

%==================================================================

\begin{references}

\bibitem{BT1986}
F.\ P.\ Bowden and D.\ Tabor,
  {\it The Friction and Lubrication of Solids}
  (Clarendon Press, Oxford, 1986).

\bibitem{P0}
B.\ N.\ J.\ Persson,
  {\it Sliding Friction: Physical Principles and Applications}
  (Springer-Verlag, Berlin, 1998).

\bibitem{P00}
B.\ N.\ J.\ Persson,
  Surf.\ Sci.\ Reports {\bf 33}, 83 (1999).

\bibitem{Mate95}
C.\ M.\ Mate,
  In: {\it Handbook of Micro/Nano Tribology},
  edited by B.\ Bhushan
  (CRC Press, Boca Raton, 1995) p.\ 167.

\bibitem{Gen}
{\it Fundamentals of Friction: Macroscopic and Microscopic Processes,}
  edited by I.\ L.\ Singer and H.\ M.\ Pollock
  (Kluwer, Dordrecht, 1992);
{\em Physics of Sliding Friction,}
  edited by B.\ N.\ J.\ Persson
  (Kluwer, Dordrecht, 1996).

\bibitem{Gen2}
G.\ Reiter, A.\ L.\ Demirel, and S.\ Granick,
  Science {\bf 263}, 1741 (1994);
B.\ Brushan, J.\ N.\ Israelachvili, and U.\ Landman,
  Nature {\bf 374}, 607 (1995);
A.\ D.\ Berman, W.\ A.\ Ducker, and J.\ N.\ Israelachvili,
  Langmuir {\bf 12}, 4559 (1996).

\bibitem{Robbins2}
M.\ O.\ Robbins,
  {\it Jamming, Friction and Unsteady Rheology},
  to appear in
  {\it Jamming and Rheology: Constrained Dynamics on
  Microscopic and Macroscopic Scales},
  edited by A.\ J.\ Liu and S.\ R.\ Nagel
  (Taylor and Francis, London, 2000 (cond-mat/9912337));
M.\ O.\ Robbins and M.\ H.\ M\"user,
  {\it Computer Simulation of Friction, Lubrication and Wear},
  to appear in
  {\it Handbook of Modern Tribology},
  edited by B.\ Bhushan
  (CRC Press, Boca Raton, 2000 (cond-mat/0001056)).

\bibitem{Gao} 
J.\ Gao, W.\ D.\ Luedtke, and U.\ Landman,
  J.\ Phys.\ Chem.\ B {\bf 101}, 4013 (1997);
  J.\ Phys.\ Chem.\ B {\bf 106}, 4309 (1997).

\bibitem{Jeam}  
J.\ E.\ Hammerberg, B.\ L.\ Holian, J.\ R\"oder, 
A.\ R.\ Bishop, and S.\ J.\ Zhou,
  Physica D {\bf 123}, 330  (1998);
R.\ P.\ Mikulla, J.\ E.\ Hammerberg, P.\ S.\ Lomdahl, and B.\ L.\ Holian, 
  Mat.\ Res.\ Soc.\ Symp.\ Proc.\ {\bf 522}, 385 (1998).

\bibitem{Persson}
B.\ N.\ J.\ Persson,
  \prl {\bf 71}, 1212 (1993);
  \prb {\bf 48}, 18140 (1993);
  J.\ Chem.\ Phys.\ {\bf 103}, 3849 (1995).

\bibitem{force}
O.\ M.\ Braun, T.\ Dauxois, M.\ Paliy, and M.\ Peyrard, 
  \prl {\bf 78}, 1295 (1997);
  Phys.\ Rev.\ E {\bf 55}, 3598 (1997);
M.\ Paliy, O.\ Braun, T.\ Dauxois, and B.\ Hu,
  Phys.\ Rev.\ E {\bf 56}, 4025 (1997).

\bibitem{BBR97}
M.\ Weiss and F.-J.\ Elmer,
  \prb {\bf 53}, 7539 (1996);
  Z.\ Phys. B {\bf 104}, 55 (1997);
O.\ M.\ Braun, A.\ R.\ Bishop, and J.\ R\"{o}der,
  \prl {\bf 79}, 3692 (1997);
O.\ M.\ Braun, B.\ Hu, A.\ Filippov, and A.\ Zeltser,
  \pre {\bf 58}, 1311 (1998).

\bibitem{Urbakh-FK}
M.\ G.\ Rozman, M.\ Urbakh, and J.\ Klafter,
  \prl {\bf 77}, 683 (1996);
  \pre {\bf 54}, 6485 (1996);
  Europhys. Lett. {\bf 39}, 183 (1997).

\bibitem{HMR1999-MR2000b-MR2000a}
G.\ He, M.\ H.\ M\"user, and M.\ O.\ Robbins,
  Science {\bf 284}, 1650 (1999);
M.\ H.\ M\"user and M.\ O.\ Robbins,
  \prb {\bf 64}, 2335 (2000);
M.\ H.\ M\"user, L.\ Wenning, and M.\ O.\ Robbins,
  \prl (2000) (cond-mat/0004494).

\bibitem{TR1990a}
P.\ A.\ Thompson and M.\ O.\ Robbins,
  Science {\bf 250}, 792 (1990);
M.\ O.\ Robbins and P.\ A.\ Thompson,
  Science {\bf 253}, 916 (1991).

\bibitem{SR1993}
M.\ J.\ Stevens and M.\ O.\ Robbins,
  \pre {\bf 48}, 3778 (1993).

\bibitem{Robbins}
P.\ A.\ Thompson, G.\ S.\ Grest, and M.\ O.\ Robbins,
  \prl {\bf 68}, 3448 (1992);
P.\ A.\ Thompson, M.\ O.\ Robbins and G.\ S.\ Grest, 
  Israel J.\ Chem.\ {\bf 35}, 93 (1995);
A.\ Baljon and M.\ O.\ Robbins, 
  MRS Bulletin {\bf 22}, 22 (1997).

\bibitem{SRC1996}
M.\ Cieplak, E.\ D.\ Smith, and M.\ O.\ Robbins,
  Science {\bf 265}, 1209 (1994);
E.\ D.\ Smith, M.\ O.\ Robbins, and M.\ Cieplak,
  \prb {\bf 54}, 8252 (1996).

\bibitem{HR2000}
G.\ He and M.\ O.\ Robbins,
  Tribology Letters (2000) (cond-mat/0008196).

\bibitem{BDP97}
O.\ M.\ Braun, T.\ Dauxois, and M.\ Peyrard,
  \prb {\bf 56}, 4987 (1997).

\bibitem{BBR99}
O.\ M.\ Braun, A.\ R.\ Bishop, and J.\ R\"{o}der,
  Phys.\ Rev.\ Lett.\ {\bf 82},  3097 (1999).

\bibitem{HS1990}
M.\ Hirano and K.\ Shinjo,
  \prb {\bf 41}, 11 837 (1990).

\bibitem{SJS1996}
M.\ R.\ S{\o}rensen, K.\ W.\ Jacobsen, and P.\ Stoltze,
  \prb {\bf 53}, 2101 (1996).

\bibitem{Plast} % Persson last squeezed layer
P.\ Ballone and B.\ N.\ J.\ Persson,
  J.\ Chem. Phys.\ (in press).

\bibitem{Krim}
J.\ Krim, D.\ H.\ Solina, and R.\ Chiarello,
  \prl {\bf 66}, 181 (1991);
J.\ Krim and R.\ Chiarello,
  J.\ Vac.\ Sci.\ Technol.\ A {\bf 9}, 2566 (1991).

\bibitem{Gardiner} 
C.\ W.\ Gardiner,
  {\it Handbook of Stochastic Methods}
  (Springer-Verlag, Berlin, 1983).

\bibitem{Braun-new} 
O.\ M.\ Braun (unpublished).

\bibitem{vibrat}
J.\ W.\ Gadzuk and A.\ C.\ Luntz,
  Surf.\ Sci.\ {\bf 144}, 429 (1984);
P.\ Avouris and B.\ N.\ J.\ Persson,
  J.\ Phys.\ Chem.\ {\bf 88}, 837 (1984);
D.\ C.\ Langreth,
  Phys.\ Scripta {\bf 35}, 185 (1987);
R.\ G.\ Tobin,
  Surf.\ Sci.\ {\bf 183}, 226 (1987).

\bibitem{B1}
O.\ M.\ Braun, A.\ I.\ Volokitin, and V.\ P.\ Zhdanov,
  Usp.\ Fiz.\ Nauk {\bf 158}, 421 (1989)
  [Sov.\ Phys.\ Usp.\ {\bf 32}, 605 (1989)].

\bibitem{B2}
O.\ M.\ Braun,
  Surface Sci.\ {\bf 213}, 336 (1989);
O.\ M.\ Braun and A.\ I.\ Volokitin,
  Fiz.\ Tverd.\ Tela {\bf 28}, 1008 (1986) % 1008-1014
  [Sov.\ Phys.\ - Solid State (USA) {\bf 28}, 564 (1986)].

\bibitem{P1} % curved algorithm
B.\ N.\ J.\ Persson, 
  \prb (in press).

\bibitem{Joanna}
A.\ R.\ Bishop, O.\ M.\ Braun, M.\ V.\ Paliy, and J.\ R\"{o}der,
  submitted to \pre (2000).

\bibitem{Sokol1}
J.\ B.\ Sokoloff,
  \prb {\bf 42}, 760, 6745(E) (1990);
%\bibitem{PN96}
B. N. J. Persson and A. Nitzan,
  Surf. Sci.\ {\bf 367}, 261 (1996);
%\bibitem{Sokol2}
J.\ B.\ Sokoloff and M.\ S.\ Tomassone,
  \prb {\bf 57}, 4888 (1998).

\bibitem{Risken}
H.\ Risken,
  {\it The Fokker-Planck Equation}
  (Springer-Verlag, Berlin, 1984).

\bibitem{P1994}
B.\ N.\ J.\ Persson,
  \prb {\bf 50}, 4771 (1994).

\bibitem{BK1967}
R.\ Burridge and L.\ Knopoff,
  Bull.\ Seismol.\ Soc.\ Am.\ {\bf 57}, 3411 (1967).

\bibitem{R1983}
J.\ H.\ Dieterich,
  J.\ Geophys.\ Res.\ {\bf 84}, 2169 (1979);
A.\ Ruina, 
  J.\ Geophys.\ Res.\ {\bf 88}, 10 359 (1983);
J.\ R.\ Rice and A.\ L.\ Ruina,
  J.\ Appl.\ Mech.\ {\bf 50}, 343 (1983);
J.\ C.\ Gu, J.\ R.\ Rice, A.\ L.\ Ruina, and S.\ T.\ Tse,
  J.\ Mech.\ Phys.\ Sol.\ {\bf 32}, 167 (1984).

\bibitem{CLS1994}
J.\ M.\ Carlson, J.\ S.\ Langer, and B.\ E.\ Shaw,
  Rev.\ Mod.\ Phys.\ {\bf 66}, 657 (1994).

\bibitem{P1995}
B.\ N.\ J.\ Persson,
  \prb {\bf 51}, 13 568 (1995).

\bibitem{P1997}
B.\ N.\ J.\ Persson,
  \prb {\bf 55}, 8004 (1997).

\end{references}
\end{document}